\documentclass[graybox]{svmult}

\usepackage{epsfig}
\usepackage{mathptmx}   
\usepackage{helvet}     
\usepackage{courier}    
\usepackage{type1cm}    
\usepackage{makeidx}    
\usepackage{graphicx}   
\usepackage{multicol}   
\usepackage[bottom]{footmisc}   

\begin{document}

\title{Non-equilibrium dynamics of quantum systems: order parameter evolution,
defect generation, and qubit transfer}

\author{Shreyoshi Mondal$^{(1)}$, Diptiman Sen$^{(2)}$ and K. 
Sengupta$^{(1,3)}$}

\institute{$^{(1)}$ Theoretical Physics Department, Indian Association for the
Cultivation of Sciences, Jadavpur, Kolkata 700 032, India \\
$^{(2)}$ Center for High Energy Physics, Indian Institute of Science, 
Bangalore 560 012, India \\
$^{(3)}$ Theoretical Condensed Matter Physics Division, Saha Institute of
Nuclear Physics, 1/AF Bidhannagar, Kolkata 700 064, India}

\titlerunning{Non-equilibrium dynamics of quantum systems $\cdots$}
\maketitle

\section{Introduction}
\label{mssintro}

The properties of systems near quantum critical points (QCPs) have
been studied extensively in recent years \cite{msssubir1,mssma}. A
QCP is a point across which the symmetry of the ground state of a
quantum system changes in a fundamental way; such a point can be
accessed by changing some parameter, say $\lambda$, in the
Hamiltonian governing the system. The change in the ground state
across a QCP is mediated by quantum fluctuations. Unlike
conventional thermal critical points, thermal fluctuations do not
play a crucial role in such transitions. Similar to its thermal
counterparts, the low energy physics near a QCP is associated with
a number of critical exponents which characterize the universality
class of such a transition. Amongst these exponents, the dynamical
critical exponent $z$ provides the signature of the relative scaling of
space and time at the transition and has no counterpart in thermal
phase transitions. The other exponent which is going to be
important for the purpose of this review is the well-known
correlation length exponent $\nu$. These exponents are formally
defined as follows. As we approach the critical point at $\lambda =
\lambda_c$, the correlation length diverges as $\xi \sim |\lambda -
\lambda_c|^{-\nu}$, while the gap between the ground state and first
excited state vanishes as $\Delta E \sim \xi^{-z} \sim |\lambda -
\lambda_c|^{z\nu}$. Exactly at the critical point $\lambda =
\lambda_c$, the energy of the low-lying excitations vanishes at some
wave number ${\vec k}_0$ as $\omega \sim |{\vec k} - {\vec k}_0|^z$.
The critical exponents are independent of the details of the
microscopic Hamiltonian; they depend only on a few parameters such
as the dimensionality of the system and the symmetry of the order
parameter. These features render the low energy equilibrium physics
of a quantum system near a QCP truly universal.

In contrast to this well-understood universality of the equilibrium
properties of a system near a QCP, relatively few universal features are
known in the non-equilibrium behavior of a quantum system. Initial studies
in this field aimed at understanding the near-equilibrium finite temperature
dynamics near a quantum critical point using the Boltzman equation approach
\cite{msskedar1}. Such a dynamics is useful in making contact with
experiments which are always carried out at finite temperature. Moreover,
the excitations near a quantum critical critical point with a non-zero
value of $\eta$ do not have a simple pole structure like that of the
conventional quasiparticle excitations of condensed matter systems;
this property makes such a dynamics interesting in its own right.

More recently, significant theoretical \cite{mssks2,msslevitov1,
mssdem1,mssanatoly0,mssks3} and experimental \cite{mssbloch1}
endeavors have focussed on out-of-equilibrium dynamics of closed
quantum critical systems. On the experimental front, it has been
possible, in ultracold atom systems, to gain unprecedented control
over the measurement of out-of-equilibrium properties of quantum
systems \cite{mssbloch1}. On the theoretical front, such studies can
be broadly classified into two distinct categories. The first type
involves a study of the time evolution of a quantum system after a
rapid quench through a quantum critical point. Such a study yields
information about the order parameter dynamics across a quantum
critical point. It turns out that such a dynamics exhibits a
universal signature of the quantum critical point crossed during the
quench. The second type involves a study of defect production during
slow non-adiabatic dynamics through a quantum critical point. Such a
defect production mechanism was first pointed out for dynamics
through thermal critical points in Refs. \cite{msski1,msszu1}. For a
slow enough quenches through quantum critical points, the density of
defects produced are known to depend on $z$ and $\nu$ which characterize 
the critical point \cite{mssanatoly1,mssmondal1,mssmondal2}.

Quantum communication in spin systems has also been a subject of
intense study recently. Following the seminal work in Ref.
\cite{mssbose1}, a tremendous amount of theoretical effort has
been put in to understand the nature of qubit transfer through one-
or multi-dimensional spin systems \cite{msscomrev1}. One of the
major goals of such studies is to characterize the fidelity of
the transfer of a qubit across such a spin system. The maximization of
both the fidelity and the speed of transfer, in moving a qubit through
a spin chain, is an issue of great interest in such studies.

In this article, we will review some studies of sudden and slow zero
temperature non-equilibrium dynamics of closed quantum systems
across critical points. In Sect. \ref{mssquench:dyn}, we consider a
sudden quench across a quantum critical point. We study the order
parameter dynamics of one-dimensional ultracold atoms in an optical
lattice in Sect. \ref{mssdipole} and of the infinite range
ferromagnetic Ising model in Sect. \ref{mssinfising}. We demonstrate
that the dynamics shows universal signatures of the QCP across which
the system is quenched. In Sect. \ref{mssnon-adiabatic}, we discuss
defect production for slow non-adiabatic dynamics; typically, we
find that the density of defects scales as an inverse power of the
quench time $\tau$, where the power depends on the dimensionality
$d$ of the system, and the exponents $z$ and $\nu$. In Sect.
\ref{msscrit_surf}, we discuss the time evolution of the system
across a quantum critical surface; we find that the defect scaling
exponent in this case depends on the dimensionality of the critical
surface. This is confirmed by a study of defect production in the
Kitaev model, which is an exactly solvable model of spin-1/2's on a
honeycomb lattice. In Sect. \ref{mssnon-linear}, we study the effect
of quenching across a QCP in a non-linear way; we find that the defect 
scaling exponent also depends on the degree of non-linearity. We illustrate 
these ideas by studying two exactly solvable spin-1/2 models in one 
dimension. In Sect. \ref{mssexpt}, we discuss a number of experimental 
systems where our results on defect scaling can possibly be checked. 
Finally, in Sect. \ref{mssquantum:comm}, we show that non-equilibrium 
dynamics, in one- and two-dimensional Heisenberg spin models, can be 
engineered to maximize the fidelity and speed of the transfer of qubits.

\section{Quench dynamics}
\label{mssquench:dyn}

\subsection {Ultracold atoms in an optical lattice}
\label{mssdipole}

In this section, we shall study a system of ultracold spinless
bosons in a one-dimensional (1D) optical lattice in the presence of a
harmonic trap potential \cite{mssks1}. We will
restrict ourselves to the Mott phase of the bosons and
will study their response to a shift in the position of the trap
potential. Such a shift acts as an effective ``electric field" for
the bosons whose Hamiltonian is given by \cite{mssks1}
\begin{equation} \mathcal{H} ~=~ -t ~\sum_{ij} \left( b_i^\dagger b_j+
b_j^\dagger b_i \right) ~+~ \frac{U}{2} ~\sum_i n_i (n_i -1) ~-~ E ~\sum_i
{\bf e} \cdot {\bf r}_i n_i, \label{mssham1d} \end{equation}
where $ij$ represents pairs of nearest neighbor
sites of the optical lattice, $n_i = b_i^\dagger b_i$ is the
number operator for the bosons, ${\bf r}_i$ are the dimensionless
spatial coordinates of the lattice sites (the lattice spacing is
unity), ${\bf e}$ is a unit vector in the direction of the applied
electric field, and the effective electric field $E$ (in units of
energy) can be deduced from the shift $a$ of the center of the trap
as $E= -a \partial V_{trap}(x)/\partial x$. In what follows, we will
restrict ourselves to $|U-E|, t \ll E,U$. We note that such
a regime has been achieved in experiments \cite{mssbloch1}.

In the presence of such an electric field, our classical intuition
suggests that all the bosons would gather in the last site of the
1D chain thereby minimizing their energy. However, this
does not happen for two reasons. First, the bosons are interacting
and a state where all the bosons are in a single site leads to a huge
interaction energy cost. But more importantly, even
non-interacting bosons (or in the parameter regime $E \gg Un_0$ for
interacting bosons) do not exhibit this behavior. To understand
this, we note that when $U=0$, $\mathcal{H}$ is simply the
Wannier-Stark Hamiltonian whose wave functions, in the limit of
strong electric fields ($t \ll E$), are well localized Bessel
functions. Thus for $E \gg t$, the bosons remain localized in their
respective lattices. It turns out that for realistic optical
lattices where interband energy spacings are large compared to both
$U$ and $E$, the Zener tunneling time, i.e., the time taken by
the bosons to reach the final ground from this metastable Mott
state, is of the order of milliseconds and is larger than the system
lifetime \cite{mssbloch1}. Our strategy will therefore be to start
from the parent Mott state of these localized bosons, identify the
complete set of states resonantly coupled to this parent state,
obtain the effective Hamiltonian within the subspace of these states, and
determine its spectrum and correlations. This effective Hamiltonian is
expected to describe the low energy behavior of the system.

\begin{figure} \centerline{\includegraphics[width=2.2in]{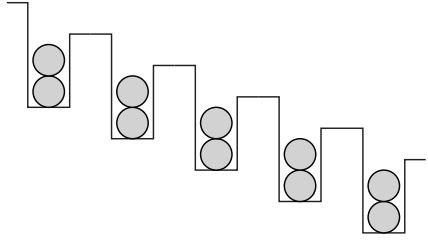}}
\caption{Schematic representation of the parent Mott insulating
state with $n_0=2$. Each well represents a local minimum of the
optical lattice potential --- we number these as 1-5 from the left. The
potential gradient leads to a uniform decrease in the on-site energy
of an atom as we move to the right. The grey circles are the $d_i$
bosons of Eq. (\ref{msshd}). The vertical direction represents
increasing energy: the repulsive interaction energy between the
atoms is realized by placing atoms vertically within each well, so
that each atom displaces the remaining atoms upwards along the
energy axis. We have chosen the diameter of the atoms to equal the
potential energy drop between neighboring wells --- this corresponds
to the condition $U=E$. Consequently, {\em a resonant transition is
one in which the top atom in a well moves horizontally to the top of
a nearest neighbor well}; motions either upwards or downwards are
non-resonant.} \label{mssfig1} \end{figure}

\begin{figure} \centerline{\includegraphics[width=2.2in]{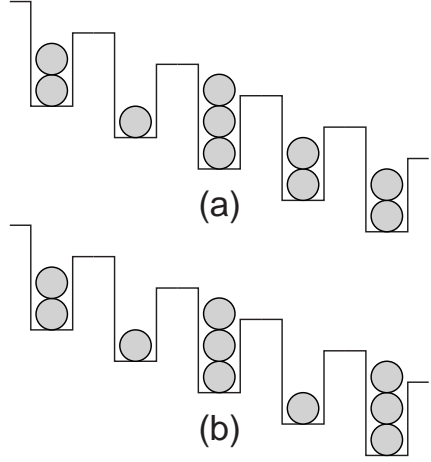}}
\caption{Notation as in Fig. \ref{mssfig1}. ({\em a}) A dipole on sites 2 and
3; this state is resonantly coupled by an infinitesimal $t$ to the Mott
insulator in ({\em a}) when $E=U$. ({\em b}) Two dipoles between sites 2 and
3 and between 4 and 5; this state is connected via multiple resonant
transitions to the Mott insulator for $E=U$.} \label{mssfig2} \end{figure}

The parent Mott state and its resonant dipole excitations are shown
in Figs. \ref{mssfig1} and \ref{mssfig2} \cite{mssks1}. A dipole here
consists of a bound pair of hole at site $i$ and an additional
particle at its neighboring $i+1$ site. We note that the dipole
excitations cost an energy $U-E$ and hence become energetically
favorable when the electric field exceeds the interaction energy.
However, once a dipole forms between two adjacent sites,
these sites cannot participate in the formation of another dipole
since the resultant state lies out of the resonant subspace
\cite{mssks1}. This leads to a constraint on the dipole number on
any given link $\ell$ connecting two sites, namely, $n_{\ell}^d \le 1$.
Similar reasoning, elaborated in Ref. \cite{mssks1}, shows that
there can be at most one dipole on two adjacent links: $n_{\ell}^d
n_{\ell+1}^d=0$. The effective Hamiltonian of these dipoles can be
written in terms of the dipole annihilation and creation operators
$d_{\ell}$ and $d_{\ell}^\dagger$ as
\begin{equation} \mathcal{H}_d ~=~ - t\sqrt{n_0 (n_0 +1)} ~\sum_{\ell}
\left( d_\ell + d_\ell^\dagger \right) ~+~ (U-E) ~\sum_\ell d^\dagger_\ell
d_\ell. \label{msshd} \end{equation}
Note that the presence of boson hopping leads to non-conservation of
the dipole number since such a hopping can spontaneously create or
destroy dipoles on a given link. Also, ${\mathcal H}_d$ needs to be
supplemented by the constraint conditions $n_{\ell}^d \le 1$ and
$n_{\ell}^d n_{\ell+1}^d =0$.

The phase diagram of the dipolar system can be easily found by
inspecting ${\mathcal H}_d$. For $(U-E)/t= \lambda \to \infty$, the
ground state of the system represents a vacuum of dipoles. In
contrast, for $\lambda \to -\infty$, the ground state is doubly degenerate
because there are two distinct states with maximal dipole number:
$(\cdots d_1^\dagger d_3^\dagger d_5^\dagger \cdots)|0 \rangle$ and $(\cdots
d_2^\dagger d_4^\dagger d_6^\dagger \cdots)|0 \rangle$. This immediately
suggests the existence of an Ising QCP at some
intermediate value of $\lambda$, associated with an order parameter
$\Delta = \sum_{\ell} (-1)^{\ell} d_\ell^\dagger d_\ell$ which is a
density wave of dipoles with a period of 2 lattice spacings. Further
analytic evidence for an Ising QCP can be
obtained by examining the excitation spectra for the limiting
$\lambda$ regimes, and noting their similarity to those on either
side of the critical point in the quantum Ising chain \cite{msssubir1}.

For $\lambda \to \infty$, the lowest excited states are
single dipoles: $|\ell \rangle = d_{\ell}^\dagger |0 \rangle$.
There are $N$ such states (where $N$ is the number of sites), and, at
$\lambda =\infty$, they are all degenerate with energy $U-E$. The
degeneracy is lifted at second order in a perturbation theory in
$1/\lambda$. By a standard approach using canonical transformations,
these corrections can be described by an effective Hamiltonian,
$\mathcal{H}_{d,{\rm eff}}$, which acts entirely within the subspace
of single dipole states. We find that
\begin{equation} \mathcal{H}_{d,{\rm eff}} = (U-E) \sum_{\ell} \Bigl[ |\ell
\rangle \langle \ell | + \frac{n_0 (n_0 + 1)}{\lambda^2} \left( |\ell\rangle
\langle \ell | + |\ell \rangle \langle \ell+1 | + | \ell+1 \rangle \langle
\ell | \right) \Bigr]. \label{msshdeff} \end{equation}
Notice that, quite remarkably, a local dipole hopping term has
appeared in the effective Hamiltonian. The constraints ($n_{\ell}^d
\le 1$ and $n_{\ell}^d n_{\ell+1}^d =0$) played a crucial role in
the derivation of (\ref{msshdeff}). Upon considering perturbations to
$|\ell \rangle$ from the first term in (\ref{msshd}), it initially seems
possible to obtain an effective matrix element between any two states $|\ell
\rangle$ and $|\ell^{\prime} \rangle$. However this connection can generally
happen via two possible intermediate states, $|\ell \rangle \to
d^\dagger_{\ell} d^\dagger_{\ell^{\prime}} | 0 \rangle \to |\ell^{\prime}
\rangle$ and $|\ell \rangle \to | 0 \rangle \to |\ell^{\prime} \rangle$, and
the contributions of the two processes exactly cancel each other for most
$\ell$, $\ell^{\prime}$. Only when the constraints block the first of these 
processes is a residual matrix element possible. It is a simple matter to
diagonalize $\mathcal{H}_{d,{\rm eff}}$ by going to momentum space;
we then find a single band of dipole states. The lowest energy dipole
state has momentum $\pi$: the softening of this state upon reducing
$\lambda$ is then consistent with the appearance of a density wave
order with period 2. The higher excited states at large $\lambda$
consist of multiparticle continua of this band of dipole states,
just as in the Ising chain \cite{msssubir1}. A related analysis can be
carried out for $\lambda \to -\infty$, and the results are
similar to those for the ordered state in the quantum Ising
chain \cite{msssubir1}. The lowest excited states form a single band of
domain walls between the two filled dipole states, and above them
are the corresponding multiparticle continua. At an intermediate critical
electric field $E_c = U + 1.310 t \sqrt{n_0 (n_0 + 1)}$, the system undergoes
a quantum phase transition lying in the Ising universality class \cite{mssks1}.

Having obtained the equilibrium phase diagram for the model, we now
consider the quench dynamics of the dipoles when the value of the
electric field is suddenly quenched \cite{mssks2}. We assume that the
atoms in the 1D lattice are initially in the ground state
$|\Psi_G\rangle $ of the dipole Hamiltonian (\ref{mssham1d}) with
$E=E_i \ll E_c$. This ground state corresponds to a dipole vacuum.
Consider shifting the center of the magnetic trap so that the new
potential gradient is $E_f$. If this change is done suddenly, the
system initially remains in the old ground state. The state of the
system at time $t$ is therefore given by
\begin{eqnarray} |\Psi(t)\rangle =\sum_n c_n \exp(-i\epsilon_n t) |n
\rangle , \end{eqnarray}
where $|n\rangle $ denotes the complete set of energy eigenstates of
the Hamiltonian ${\mathcal H}_d$ with $E=E_f$, $\epsilon_n =
\,\,\langle n|{\mathcal H}_d [E_f]|n\rangle $ is the energy
eigenvalue corresponding to state $|n\rangle $, and $c_n =\,\,\langle n|
\Psi(t=0)\rangle = \langle n|\Psi_G\rangle $ denotes the overlap of the old
ground state with the state $|n\rangle $. (We have set $\hbar =1$).
Notice that the state $|\Psi(t)\rangle $ is no longer the ground
state of the new Hamiltonian. Furthermore, in the absence of any
dissipative mechanism, which is the case for ultracold atoms in
optical lattices, $|\Psi(t)\rangle $ will never reach the ground
state of the new Hamiltonian. Rather, in general, we expect the
system to thermalize at long enough times, so that the correlations
are similar to those of $H_{1D} [E_f]$ at some finite temperature.

We are now in a position to study the dynamics of the Ising density
wave order parameter
\begin{equation} O = \frac{1}{N}\langle \Psi|\Delta|\Psi\rangle,
\label{mssisingO} \end{equation}
where $N$ is the number of sites. The time evolution of $O$ is given by
\begin{equation} O(t) ~=~ \frac{1}{N} ~\sum_{m,n} ~c_m c_n ~\cos\left[\left(
E_m-E_n \right)t \right] ~\langle m|\sum_\ell (-1)^\ell d_\ell^\dagger
d_\ell|n\rangle . \label{mssising} \end{equation}
Eq. (\ref{mssising}) is solved numerically using exact diagonalization
to obtain the eigenstates and eigenvalues of the Hamiltonian $H_{\rm
1D}[E_f]$. Before resorting to numerics, it is useful to
discuss the behavior of $O(t)$ qualitatively. We note that if $E_f$
is close to $E_i$, the old ground state will have a large overlap
with the new one, i.e., $c_m \sim \delta_{m1}$. Hence in this case
we expect $O(t)$ to have small oscillations about $O(t=0)$. On the
other hand, if $E_f \gg E_c$, the two ground states will have very
little overlap, and we again expect $O(t)$ to have a small
oscillation amplitude. This situation is in stark contrast with the
adiabatic turning on of the potential gradient, where the systems
always remain in the ground state of the new Hamiltonian $H_{\rm
1D}[E_f]$, and therefore has a maximal value of $\langle O\rangle$
for $E_f \gg E_c$. In between, for $E_f \sim E_c$, the ground state
$|\Psi\rangle $ has a finite overlap with many states $|m\rangle $,
and hence we expect $O(t)$ to display significant oscillations.
Furthermore, if the symmetry between the two Ising ordered states is
broken slightly (as is the case in our studies below), the time
averaged value of $O(t)$ will be non-zero.

\begin{figure} \centerline{\epsfig{file=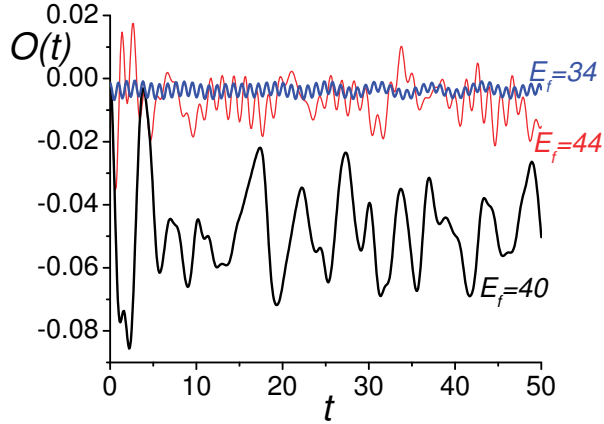,width=8cm,angle=0}}
\caption{Evolution of the Ising order parameter in (\ref{mssisingO})
under the Hamiltonian $H_{1D} [E_f]$ for $n_0 = 1$. The initial state is
the ground state of $H_{1D} [E_i]$. All the plots in this section have
$U=40$, $t=1$, and $E_i=32$, and consequently the equilibrium QCP is at
$E_c = 41.85$.} \label{mssfig3} \end{figure}

\begin{figure} \centerline{\epsfig{file=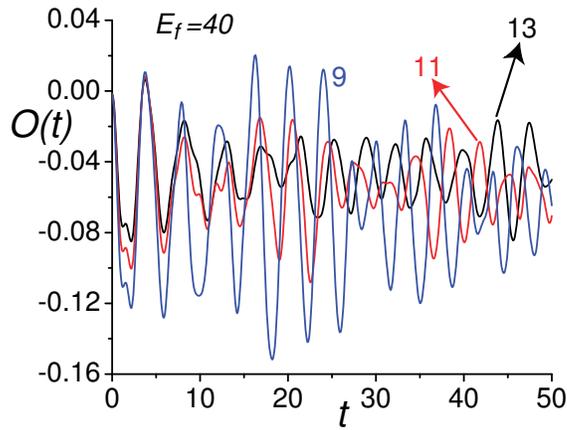,width=8cm,angle=0}}
\caption{System size ($N$) dependence of the results of Fig. \ref{mssfig3} for
$E_f=40$. The curves are labeled by the value of $N$.} \label{mssfig4}
\end{figure}

\begin{figure} \centerline{\epsfig{file=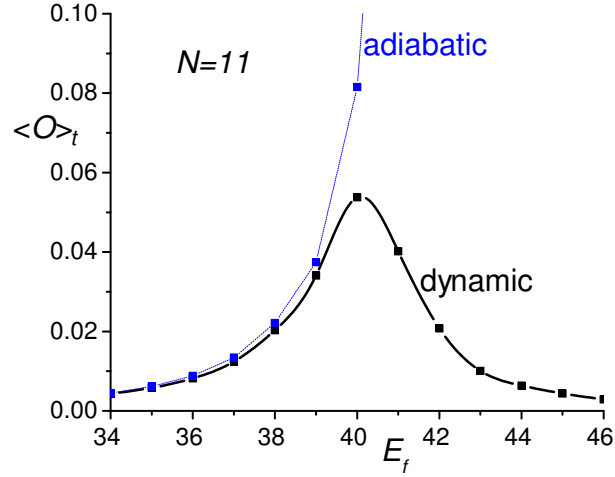,width=8cm,angle=0}}
\caption{The curve labeled `dynamic' is the long-time limit $\langle O
\rangle_t$ of the Ising order parameter in (\ref{mssising}) as a function
of $E_f$ (for $N=11$), with other parameters as in Fig. \ref{mssfig1}.
This long-time limit can be obtained simply by setting $m=n$ in
(\ref{mssising}). For comparison, in the curve labeled `adiabatic', we
show the expectation value of the Ising order $O$ in the ground state of
$H_{1D} [E_f]$; such an order would be observed if the value of $E$ was
changed adiabatically. Note that the dynamic curve has its maximal value near
(but not exactly at) the equilibrium QCP $E_c = 41.85$, where the system is
able to respond most easily to the change in the value of $E$;
this dynamic curve is our theory of the `resonant' response in the experiments
of Ref. \cite{mssbloch1} discussed in Sect. \ref{mssintro}. In contrast,
the adiabatic result {\em increases monotonically} with $E_f$ into the $E>E_c$
phase where the Ising symmetry is spontaneously broken.} \label{mssfig5}
\end{figure}

\begin{figure} \centerline{\epsfig{file=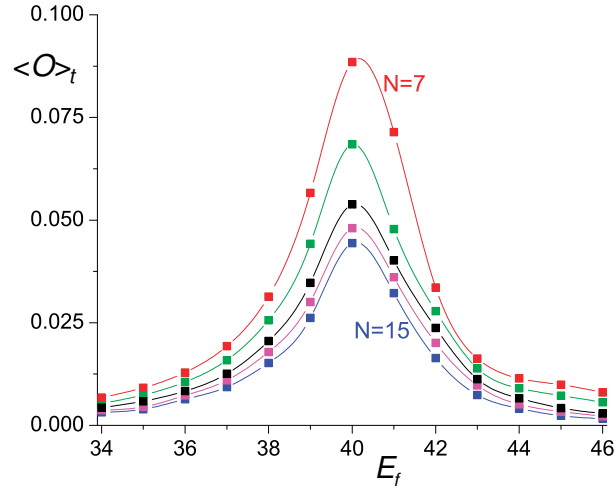,width=8cm,angle=0}}
\caption{Size dependence of the `dynamic' results in Fig. \ref{mssfig5}. The
sizes range from $N=7$ to $N=15$ (as labeled), with the intermediate values
$N=9,11,13$: $\langle O\rangle_t$ decreases monotonically with $N$.}
\label{mssfig6} \end{figure}

This qualitative discussion is supported by numerical calculations
for finite size systems with size $N=9,11,13$. For numerical
computations with finite systems, we choose systems with an odd
number of sites and open boundary conditions, so that dipole
formation on odd sites is favored, thus breaking the $Z_2$ symmetry.
The results are shown in Figs. \ref{mssfig3} - \ref{mssfig6}.

Figure \ref{mssfig3} shows the oscillations of the order parameter $O(t)$
for different values of $E_f$ for $N=13$. In agreement with our
qualitative expectations, the oscillations have maximum amplitude
when $E_f/t \approx 40$ is near the critical value $E_c/t = 41.85$.
For either $E_f \ll E_c$ or $E_f \gg E_c$, the oscillations have a
small amplitude around $O(t=0)$. Furthermore, it is only for $E_f
\approx E_c$ that the time averaged value of $O(t)$ is appreciable.
Figure \ref{mssfig4} shows the system size dependence of the time
evolution for $E_f = U=40 t$. We find that the oscillations remain
visible as we go to higher system sizes, although they do weaken
somewhat. More significantly, the time averaged value of $O(t)$
remains non-zero, and has a weaker decrease with system size. In
Fig. \ref{mssfig5}, we plot the long-time limit of the Ising order
parameter, $\langle O\rangle_t$, as a function of $E_f$, and
compare it with $O_{\rm ad}$, the value of the order parameter
when $E$ reaches $E_f$ adiabatically and the wave function is that of
the ground state at $E=E_f$. We find that $\langle O\rangle_t$
stays close to $O_{\rm ad}$ as long as there is a large overlap
between the old and the new ground states. However, as we approach
the adiabatic phase transition point, this overlap decreases and
$\langle O\rangle_t$ cannot follow $O_{\rm ad}$ any more. The
deviation of $\langle O\rangle_t$ is therefore a signature that the
system is now in a different phase for the new value of the electric
field. The `dynamic' curve in Fig. \ref{mssfig5} shows that the Mott
insulator has a resonantly strong response to an electric field
$E\sim U$ induced by the proximity of a QCP.

We comment briefly on the nature of the thermodynamic limit, $N
\to \infty$ for the results in Figs. \ref{mssfig2} and \ref{mssfig3}.
For $O_{\rm ad}$ it is clear that there is a non-zero limit only for
$E > E_c$, when it equals the order parameter of the spontaneously
broken Ising symmetry. If we assume that the system thermalizes at
long times for the dynamic case, then $\langle O \rangle_t$
corresponds to the expectation value of the equilibrium order
parameter in $H_{1D} [E_f]$ at some finite temperature. In one
dimension, it is not possible to break a discrete symmetry at finite
temperatures, and so the thermodynamic limit of the order parameter must
always vanish. By this reasoning, we expect $\langle O \rangle_t$ to also
vanish in the thermodynamic limit. This is consistent with the results in
Fig. \ref{mssfig6}, where we show the $N$ dependence of the long-time limit
$\langle O \rangle_t$. Our data are at present not extensive enough to
definitely characterize the dependence of $\langle O \rangle_t$ on $N$.

\subsection{Infinite range Ising model in a transverse field}
\label{mssinfising}

The analysis of the quench dynamics of 1D ultracold atoms do not permit an
analytical description of the long-time value of the order parameter. In
particular, the system size dependence of the peak height of $\langle O
\rangle_t$ is not easy to understand analytically in this model. For this 
purpose, we now consider a
simple model system, the infinite range ferromagnetic spin-1/2 Ising model in
a transverse field, and study its quench dynamics due to a sudden variation of
the transverse field. The model Hamiltonian is given by \cite{mssks3}
\begin{equation} H ~=~ - ~\frac{J}{N} ~\sum_{i<j} ~S_i^z S_j^z ~-~ \Gamma ~
\sum_i ~S_i^x ~, \label{mssham1} \end{equation}
where $S^a_i = \sigma^a_i /2$, $a = x, y, z$, denote
the components of the spin-1/2 operator represented by the standard
Pauli spin matrices $\sigma^a$. Here we assume that $J \ge 0$
(ferromagnetic Ising interaction). This Hamiltonian is invariant
under the $Z_2$ symmetry $S_i^x \to S_i^x$, $S_i^y \to - S_i^y$, and
$S_i^z \to - S_i^z$. (The $Z_2$ symmetry would not be present if
there was a longitudinal magnetic field coupling to $\sum_i S_i^z$).
We take $\Gamma \ge 0$ without loss of generality since we can
always resort to the unitary transformation $S_i^x \to - S_i^x$,
$S_i^y \to - S_i^y$ and $S_i^z \to S_i^z$, which flips the sign of
$\Gamma$ but leaves $J$ unchanged. Eq. (\ref{mssham1}) can be written as
\begin{eqnarray} H &=& -~ \frac{J}{2N} ~(~ S_{tot}^z ~)^2 ~-~ \Gamma ~
S_{tot}^x ~, \label{mssham2} \\
{\rm where} ~~~ S_{tot}^z &=& ~\sum_i ~S_i^z, ~~~~~S_{tot}^x ~=~ \sum_i ~
S_i^x , \end{eqnarray}
and we have dropped a constant $(J/2N) \sum_i (S_i^z)^2 = J/8$ from the
Hamiltonian in Eq. (\ref{mssham2}). This model has been studied extensively,
particularly from the point of view of quantum entanglement \cite{mssvidal1}.
Note that this model differs from the one studied in Ref. 
\cite{msschakrabarti2}, where the spins were taken to be living on two
sub-lattices, with Ising interactions only between spins on different
sub-lattices.

We begin with a mean field analysis of the thermodynamics of the model
described by Eq. (\ref{mssham1}). Denoting the mean field value $m = \sum_i
\left<S_i^z \right>/N$, the Hamiltonian governing any one of the spins is
given by
\begin{equation} h ~=~ - J m S^z_{tot} ~-~ \Gamma S^x_{tot} ~. \end{equation}
This is a two-state problem whose partition
function can be found at any temperature $T$. If $\beta = 1/(k_B T)$,
we find that $m$ must satisfy the self-consistent equation
\begin{equation} m ~=~ \frac{Jm}{2 \sqrt{\Gamma^2 + J^2 m^2}} ~\tanh \left(
\frac{\beta \sqrt{ \Gamma^2 + J^2 m^2}}{2} \right) ~. \label{mssmag}
\end{equation}
This always has the trivial solution $m=0$. In the limit of zero
temperature, there is a non-trivial solution if $\Gamma < J/2$, with
$|m| = (1/2) \sqrt{1 - 4 \Gamma^2 /J^2}$; the energy gap in that
case is given by $J/2$. If $\Gamma > J/2$, we have $m=0$ and the gap
is given by $\Gamma - J/2$. Hence there is a zero temperature phase
transition at $\Gamma_c = J/2$. The $Z_2$ symmetry mentioned after
Eq. (\ref{mssham1}) is spontaneously broken and $<S_i^z>$ becomes
non-zero when one crosses from the paramagnetic phase at $\Gamma >
J/2$ into the ferromagnetic phase $\Gamma < J/2$.

\begin{figure}[htb] \centerline{\epsfig{figure=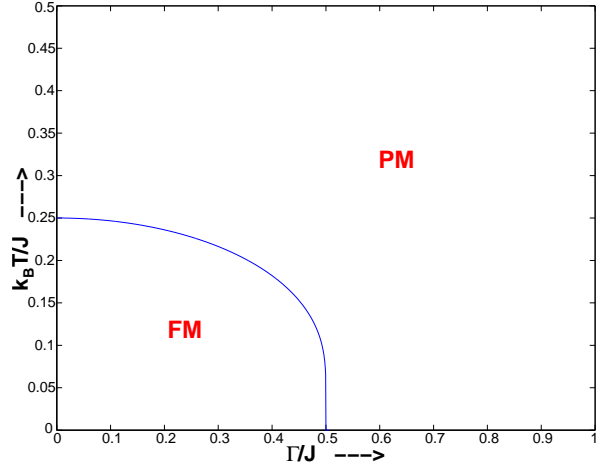,width=8.5cm}}
\caption{Phase diagram of the model in mean field theory. FM and PM denote
ferromagnetic and paramagnetic regions respectively.} \label{mssfig7}
\end{figure}

In the plane of $(k_B T / J , \Gamma /J)$, there is a ferromagnetic (FM)
region in which the solution with $m \ne 0$ has a lower free energy (the $Z_2$
symmetry is broken), and a paramagnetic (PM) region in which $m=0$. The
boundary between the two is obtained by taking the limit $m \to 0$ in
(\ref{mssmag}). This gives $2 \Gamma /J = \tanh (\beta \Gamma /2)$, i.e.,
\begin{equation} \frac{k_B T}{J} ~=~ \frac{\Gamma}{J} ~\left[ \ln \left(
\frac{1 + 2 \Gamma /J}{1 - 2 \Gamma /J} \right) \right]^{-1}. \end{equation}
The mean-field phase diagram is shown in Fig. \ref{mssfig7}. We note that the
exact excitation spectrum of this model can also be obtained analytically
\cite{mssks3}.

\begin{figure}[htb] \centerline{\epsfig{figure=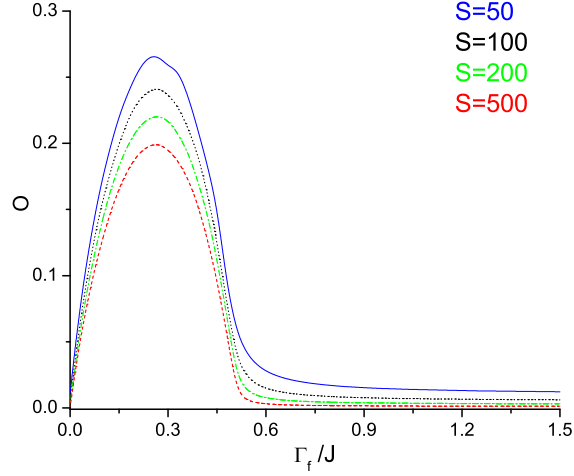,width=8.5cm}}
\caption{Dynamics of $\left<(S_{tot}^z)^2\right>/S^2$ for $S=100$ after
quenching the transverse field to different values $\Gamma_f/J$ from an initial
field $\Gamma_i/J=2$. The oscillation amplitudes are small, as seen from the
solid (red) and dotted (blue) curves corresponding to $\Gamma_f/J=0.9$ and
$0.01$ respectively, far away from the critical point $\Gamma_c/J=0.5$. The
oscillation is large in the ordered phase near the critical point as seen
from the dashed (black) curve $\Gamma_f/J=0.4$.} \label{mssfig8} \end{figure}

Having obtained the phase diagram, we now study the quench dynamics
across the QCP. To begin with, we study the
dynamics of the equal-time order parameter correlation function
(EOC) (defined as $\left<(S_{tot}^z)^2\right> /S^2$) by changing the
transverse field $\Gamma$ from an initial value $\Gamma_i \gg
\Gamma_c$ to a final value $\Gamma_f$ suddenly, so that the ground
state of the system has no time to change during the quench. In this
case, just after the quench, the ground state of the system can be expressed,
in terms of the eigenstates $\left|n\right>$ of the new Hamiltonian
${\mathcal H}_f = -(J/4S) (S_{tot}^z)^2 - \Gamma_f S_{tot}^x$ as
\begin{eqnarray} \left|\psi \right> &=& \sum_n c_n \left|n\right>,
\label{msswaveini} \end{eqnarray}
where $c_n$ denotes the overlap of the
eigenstate $\left|n\right>$ with the old ground state $\left|\psi
\right>$. As the state of the system evolves, it is given at time $t$ by
\begin{eqnarray} \left|\psi(t) \right> &=& \sum_n c_n e^{-iE_n t}
\left|n \right>, \label{msswavedy} \end{eqnarray}
where $E_n = \left<n\right|{\mathcal H}_f \left|n\right>$ are the energy
eigenvalues of the Hamiltonian ${\mathcal H}_f$. The EOC can thus be written as
\begin{equation} \left<\psi(t) \right|(S_{tot}^z)^2 /S^2 \left|\psi(t)\right> ~
=~ \sum_{m,n} ~c_n c_m ~\cos \left[\left(E_n-E_m \right)t \right] ~
\left< m\right| (S_{tot}^z)^2 /S^2 \left| n \right>. \label{mssopcorr}
\end{equation}
Eq. (\ref{mssopcorr}) can be solved numerically to obtain the time evolution 
of the EOC. We note that, similar to the case of the dipole model discussed 
in Sect. \ref{mssdipole}, we expect the amplitude of oscillations to be 
maximum when $\Gamma_f$ is near $\Gamma_c$. This is verified in Fig. 
\ref{mssfig8}. Here, we have quenched the transverse fields to $\Gamma_f/J=
0.9,0.01,\,\, {\rm and}\,\, 0.4$ starting from $\Gamma_i/J=2.0$. The 
oscillation amplitudes of the EOC for $S=100$, as shown in Fig. \ref{mssfig8},
are small for $\Gamma_f=0.9\,\, {\rm and}\,\, 0.01$, whereas it is
substantially larger for $\Gamma_f=0.4$.

\begin{figure}[htb] \centerline{\epsfig{figure=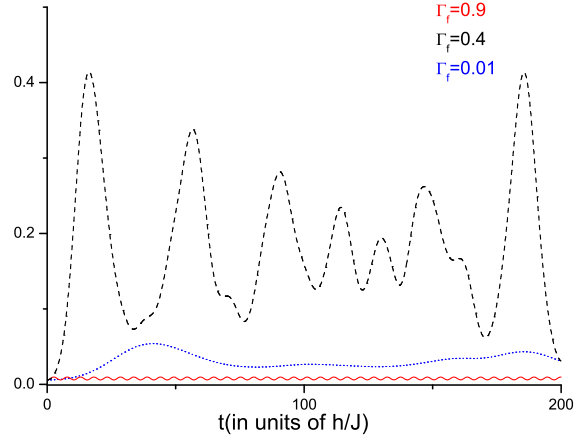,width=8.5cm}}
\caption{Plot of the long-time average $O$ as a function of $\Gamma_f/J$ for
different $S$. The solid (blue), dotted (black), dash-dotted
(green) and the dashed (red) lines represent respectively the results for
$S = 50$, $S = 100$, $S = 200$ and $S = 500$. $O$ peaks around $\Gamma_f/J=
0.25$, and the peak value decreases with increasing $S$. We have chosen
$\Gamma_i/J=2$ for all the plots.} \label{mssfig9} \end{figure}

Next, to understand the dynamics of the EOC in a little more detail,
we study its long-time averaged value given by
\begin{eqnarray} O &=& \lim_{T\to \infty} \left< \left< (S_{tot}^z)^2(t)
\right > \right>_T / S^2 \nonumber \\
&=& \frac{1}{S^2} \sum_n c_n^2 \left<n\right|(S_{tot}^z)^2 \left|n \right>
\label{msslongtime} \end{eqnarray}
for different $\Gamma_f$. Note that the long-time average depends on the
product of the overlap of the state $\left|n\right>$ with the old ground
state and the expectation of $(S_{tot}^z)^2$ in that state. From our earlier
discussion in Sect. \ref{mssdipole}, we therefore expect $O$ to have a peak
somewhere near the critical point where such an overlap is
maximized. This is verified by explicit numerical computation of Eq.
(\ref{msslongtime}) in Fig. \ref{mssfig9} for several values of $S$ and
$\Gamma_i/J=2$. We find that $O$ peaks around $\Gamma_f/J=0.25$, and
the peak height decreases slowly with increasing $S$.

To understand the position and the system size dependence of the
peak in $O$, we now look at the thermodynamic (large system size)
limit; in the present model, this is also the large $S$ and
therefore classical limit. With this observation, we study the
classical equations of motion for ${\bf S} = S\left(\cos \phi \sin
\theta, \sin \phi \sin \theta, \cos \theta \right)$ for
$\Gamma=\Gamma_f$. In the present model, $S$ is a constant. Thus in
the classical limit, we need to study the equations of motion for
$\theta$ and $\phi$. To this end, we note that the classical Lagrangian can
be written in terms of $\theta$ and $\phi$ as \cite{mssfradkin}
\begin{eqnarray} L &=& - S ~\left[ 1-\cos \theta \right] ~\frac{d \phi}{dt} ~
-~ {\mathcal H}\left[\theta,\phi \right]. \end{eqnarray}
This gives the equations of motion
\begin{eqnarray} \frac{d \theta}{dt} &=& \Gamma_f ~\sin \phi ~, \nonumber \\
\frac{d \phi}{dt} &=& -~\frac{J}{2} ~\cos \theta ~+~ \Gamma_f ~\cot
\theta \cos \phi ~. \label{msseom1} \end{eqnarray}
Eq. (\ref{msseom1}) has to be supplemented with the initial condition that
$S_{tot}^x=S$ at $t=0$. The condition $S_{tot}^x=S$ corresponds to $\theta=
\pi/2,\,\,\phi=0$ which is also a fixed point of (\ref{msseom1}). Therefore
we shall start from an initial condition which is very close to the fixed
point: $\theta=\pi/2 -\epsilon, \,\,\phi=\epsilon$, where $\epsilon$ is an
arbitrarily small constant. Further, since the motion occurs on a constant
energy surface after the quench has taken place, we have
\begin{eqnarray} \Gamma_f &=& \frac{J}{4} ~\cos^2 \theta ~+~ \Gamma_f ~ \sin
\theta \cos \phi ~. \label{mssec1} \end{eqnarray}
Using (\ref{msseom1}) and (\ref{mssec1}), we get an equation of motion for
$\theta$ in closed form,
\begin{eqnarray} \frac{d \theta}{dt} =~ \frac{\sqrt{\Gamma_f^2 \sin^2 \left(
\theta\right) -\left[ \Gamma_f -\frac{J}{4} \cos^2 \theta \right]^2}}{\sin
\theta} ~ \equiv f\left(\theta\right). \label{msseom2} \end{eqnarray}
It can be seen that the motion of $\theta$ is oscillatory and has classical
turning points at $\theta_1 = \sin^{-1} \left(\left| 1-4\Gamma_f /J
\right|\right)$ and $\theta_2 = \pi/2$. One can now obtain
$\left<(S_{tot}^z)^2\right>_T = \left<\cos^2 \theta \right>_T$ from
(\ref{msseom2}),
\begin{eqnarray} \left<\cos^2 \theta \right>_T &=& {\mathcal N} /
{\mathcal D} ~, \label{mssav1} \\
{\rm where} ~~~~{\mathcal N} &=& \int_{\theta_1}^{\theta_2} d\theta ~
\frac{\cos^2 \theta}{f\left(\theta\right)} ~=~ 4 \sqrt{8 \Gamma_f \left(
J-2 \Gamma_f \right)}/J ~, \label{mssnum1} \\
{\rm and} ~~~~~{\mathcal D} &=& \int_{\theta_1}^{\theta_2} d\theta ~
\frac{1}{f \left(\theta\right)} ~. \label{mssdeno1} \end{eqnarray}
When trying to evaluate ${\mathcal D}$, we find that the integral has an
end-point singularity at $\theta_2$; this can be regulated by a cut-off $\eta$
so that $\theta_2 = \pi/2 -\eta$. With this regularization,
${\mathcal D} = -J \ln(\eta)/\sqrt{\Gamma_f
\left(J-2\Gamma_f\right)/2}$. The cut-off used here has a physical
meaning and is not arbitrary. To see this, note that the angles
$(\theta ,\phi)$ define the surface of a unit sphere of area $4\pi$.
This surface, for a system with spin $S$, is also the phase space
which has $2S+1$ quantum mechanical states. For large $S$, the area
of the surface occupied by each quantum mechanical state is
therefore $4\pi /(2S+1) \simeq 2\pi/S$. In other words, each quantum
mechanical state will have a linear dimension of order $1/\sqrt{S}$;
this is how close we can get to a given point on the surface of the
sphere. Note that this closeness is determined purely by quantum
fluctuation and vanishes for $S\to \infty$. Thus $\eta$, which is also a
measure of how close to the point $\theta=\pi/2$ we can get, must be of the
order of $1/\sqrt{S}$; this determines the system-size dependence of
$\left<\cos^2 \theta \right>_T$. Using (\ref{mssav1}), we finally get
\begin{eqnarray} \left<\cos^2 \theta \right>_T &=& \frac{16 \Gamma_f \left(J
-2 \Gamma_f\right)}{J^2 \ln(S)}. \label{msscr1} \end{eqnarray}
Eq. (\ref{msscr1}) is one of the main results of this
section. It demonstrates that the long-time average of the EOC must be
peaked at $\Gamma_f/J=0.25$ which agrees perfectly with the exact
quantum mechanical numerical analysis leading to Fig. \ref{mssfig9}.
Moreover, it provides an analytical understanding of the $S$ (and
hence system size) dependence of the peak values of $\Gamma_f/J$; 
we conclude that the peak 
in $O$ vanishes logarithmically with the system size $S$. Such a slow 
variation with $S$ shows that it might be experimentally possible to observe 
an experimental signature of a QCP for a possible realization of this model 
with ultracold atoms where $N \sim 10^5-10^6$ \cite{mssks3}.

The results obtained in this section can be tested in two kinds of
experimental systems. One class of systems are those with long-range
dipole-dipole interactions such as KH$_2$PO$_4$ or Dy(C$_2$H$_5$SO$_4$)$_3
$9H$_2$O \cite{msschakrabarti1} which exhibit order-disorder transitions
driven by tunneling fields. The other class of systems are two-component
Bose-Einstein condensates where the inter-species interaction is
strong compared to the intra-species interaction; the relative strengths of
these interactions can be changed by tuning the system to be near a Feshbach
resonance as discussed for the $^{41}K-^{87}Rb$ system in Refs.
\cite{msscoldref1,mssbecth,mssbecexp,mssbec3}. The quench dynamics that we
have discussed can be realized by applying a radio frequency pulse to the
system and suddenly changing the frequency of the pulse.

We end this section with the observation that the resonant response of the
order parameter during quench dynamics has been found in two very disparate
models (1D ultracold atoms and infinite dimensional Ising ferromagnet) and
therefore seems to be a universal signature of the QCP through which the
system passes during its evolution.

\section{Non-adiabatic dynamics}
\label{mssnon-adiabatic}

In recent years, there have been extensive studies of what happens
when a parameter $\lambda$ in the Hamiltonian of a quantum system is
varied in time slowly (non-adiabatically) so as to take the system
through a QCP. A quantum phase transition is necessarily accompanied
by diverging length and and time scales, or, equivalently, a
vanishing energy gap between the ground state and the first excited
state \cite{msssubir1}. A consequence of this is that the system
fails to be in the adiabatic limit when it crosses a critical point.
Namely, when $\lambda$ is varied across the QCP located at $\lambda
= \lambda_c$ at a finite rate given by $1/\tau$ (where $\tau$ will
be called the quench time), the system fails to follow the
instantaneous ground state in a finite region around $\lambda_c$. As
a result, defects are produced \cite{msski1,mssdamski1}. For a
slow quench (for instance, for values of $\tau$ much larger than the
inverse band width) which takes the system across a QCP in a linear
way, it is well-known that the density of defects $n$ scales as a
power of the quench time, $n \sim 1/\tau^{d\nu/ (z \nu +1)}$, where
$\nu$ and $z$ are respectively the correlation length and the
dynamical critical exponents characterizing the critical point
\cite{mssanatoly1,mssanatoly2}. A theoretical study of a quench
dynamics requires a knowledge of the excited states of the system.
Hence, early studies of the quench problem were mostly restricted to
quantum phase transitions in exactly solvable models such as the 1D
Ising model in a transverse field \cite{mssks2,mssdziar1,msscardy},
the 1D $XY$ spin-1/2 model \cite{msslevitov,msssen1}, quantum spin
chains \cite{mssdamski2,msscaneva,msszurek}, the Bose-Hubbard model
\cite{msskollath}, the Falicov-Kimball model \cite{msseckstein} and
1D spinless fermionic chains \cite{mssmanmana}. Experimentally,
trapped ultracold atoms in optical lattices provide possibilities of
realization of many of the above-mentioned systems
\cite{mssbloch2,mssduan}. Experimental studies of defect production
due to quenching of the magnetic field in a spin-1 Bose condensate
have also been undertaken \cite{msssadler}.

A class of models in which the above power law scaling can be derived easily
is one in which, due to the existence of a mapping to a system of
non-interacting fermions, the system decomposes into a product of
two-level systems. For instance, this occurs in the 1D $XY$ spin-1/2 model
and in the two-dimensional (2D) spin-1/2 Kitaev model. In both these cases,
it turns out that the Hamiltonian is given by a sum of terms of the form
\begin{equation} H_{\vec k} ~=~ \alpha ({\vec k}) (c_{\vec k}^\dagger
c_{\vec k} ~+~ c_{-\vec k}^\dagger c_{\vec k}) ~+~ \Delta^* ({\vec k})
c_{\vec k}^\dagger c_{-\vec k}^\dagger ~+~ \Delta ({\vec k}) c_{-\vec k}
c_{\vec k}, \label{mssHk} \end{equation}
where ${\vec k}$ runs over {\it half} the Brillouin zone (BZ), and $\alpha
({\vec k})$ is real. This Hamiltonian acts on a space spanned by four states,
namely, the empty state $|0 \rangle$, two one-fermion states
$c_{\vec k}^\dagger |0 \rangle = | {\vec k} \rangle$ and
$c_{-\vec k}^\dagger |0 \rangle = | -{\vec k} \rangle$, and a two-fermion
state $c_{\vec k}^\dagger c_{-{\vec k}}^\dagger |0 \rangle = | {\vec k},
-{\vec k} \rangle$. Both the one-particle states are eigenstates of
$H_{\vec k}$ with the same eigenvalue $\alpha ({\vec k})$. On the other
hand, the states $|0 \rangle$ and $| {\vec k}, -{\vec k} \rangle$ are governed
by a $2 \times 2$ Hamiltonian given by
\begin{equation} h_{\vec k} ~=~ \left( \begin{array}{cc}
0 & \Delta ({\vec k}) \\
\Delta^* ({\vec k}) & 2 \alpha ({\vec k}) \end{array} \right). \label{msshk1}
\end{equation}
The eigenvalues of this are given by $\alpha ({\vec k}) \pm \sqrt{\alpha^2
({\vec k}) + |\Delta ({\vec k})|^2}$; since the lower eigenvalue is less than
$\alpha ({\vec k})$, the ground state lies within the subspace of the states
$|0 \rangle$ and $| {\vec k}, -{\vec k} \rangle$. We observe that the
Hamiltonian in (\ref{mssHk}) does not mix the states $|0 \rangle$ and
$| {\vec k}, -{\vec k} \rangle$ with the
states $| {\vec k} \rangle$ and $| -{\vec k} \rangle$, {\it even if}
$\alpha ({\vec k})$ and $\Delta ({\vec k})$ change with time.
Hence, if we start at time $t \to -\infty$ with a linear superposition of
$|0 \rangle$ and $| {\vec k}, -{\vec k} \rangle$, we will end at $t \to
\infty$ with a superposition of the same two states. In that case, it
is sufficient to restrict our attention to the Hamiltonian for a two-level
system given in (\ref{msshk1}). We can rewrite that as
\begin{equation} h_{\vec k} ~=~ \alpha ({\vec k}) ~I_2 ~-~ \alpha ({\vec k})
~\sigma^3_{\vec k} ~+~ \Delta ({\vec k}) ~\sigma^+_{\vec k} ~+~ \Delta^*
({\vec k}) ~ \sigma^-_{\vec k}, \label{msshk2} \end{equation}
where $I_2$ is the identity matrix, and $\sigma^3$ and $\sigma^\pm =
(\sigma^1 \pm i \sigma^2)/2$ denote the Pauli matrices. We can ignore
the term $\alpha ({\vec k}) I_2$ in (\ref{msshk2}) since this only
affects the wave function by a time-dependent phase factor.

Let us now consider what happens if $\alpha ({\vec k})$ varies linearly with
time. Then the total Hamiltonian $H_d$ for all the two-level systems can be
written as
\begin{equation} H_d ~=~ \sum_{\vec k} ~h_{\vec k}, ~~~{\rm where} ~~~
h_{\vec k} ~=~ \frac{t}{\tau} ~\epsilon ({\vec k}) ~\sigma^3_{\vec k} ~+~
\Delta({\vec k}) ~\sigma^+_{\vec k} ~+~ \Delta^* ({\vec k}) ~
\sigma^-_{\vec k} , \label{msshfd} \end{equation}
where $d$ is the number of spatial dimensions, and the sum over ${\vec k}$
runs over half the BZ. If $\epsilon ({\vec k}) > 0$, the ground state is
given by $|0 \rangle$ as $t \to -\infty$ and by $| {\vec k}, -{\vec k}
\rangle$ as $t \to \infty$. If we begin with the state $|0 \rangle$ at
$t = - \infty$ and evolve the system using the
time-dependent Schr\"odinger equation, we end at $t = \infty$ in a state
which is a superposition of states $|0 \rangle$ and $| {\vec k}, -{\vec k}
\rangle$ with probabilities $p_{\vec k}$ and $1 - p_{\vec k}$, where
$p_{\vec k}$ is given by the Landau-Zener expression \cite{msslz}
\begin{equation} p_{\vec k} ~=~ e^{-\pi \tau |\Delta({\vec k})|^2 /\epsilon
({\vec k})}. \label{msspk} \end{equation}
Note that $p_{\vec k} \to 1$ for $\tau \to 0$ (sudden quench) and $\to 0$
for $\tau \to \infty$ (adiabatic quench), as expected. Let us now assume that
the gap function $\Delta({\vec k})$ vanishes at some point ${\vec k}_0$ in
the BZ as $|\Delta({\vec k})| \simeq a_0 |{\vec k} - {\vec k}_0|^z$, while
$\epsilon ({\vec k}_0) = b_0$ is finite;
this corresponds to a QCP with an arbitrary value of $z$, but with $z \nu =
1$. The density of defects $n$ in the final state is given by the density of
fermions $n = \int d^d {\vec k} p_{\vec k}$. In the adiabatic limit $\tau
\to \infty$, this is given by
\begin{equation} \int_{{\vec k} \sim {\vec k}_0} d^d k ~e^{-\pi \tau a_0^2 |
{\vec k} - {\vec k}_0|^{2z} /b_0}~ \sim ~ 1/\tau^{d/(2z)}, \label{mssscaling}
\end{equation}
which is the expected result for $z \nu = 1$.

It is useful to note that the derivation of the scaling law in
(\ref{mssscaling}) does not require a knowledge of the precise functional form
given in (\ref{msspk}). It is enough to know that $p_{\vec k}$ must be a 
function of the form $f(\tau |\Delta ({\vec k})|^2/\epsilon ({\vec k}))$, 
where $f(x) \to 1$ for $\tau \to 0$ and $\to 0$ for $\tau \to \infty$; these 
limiting values follow from general properties of the time-dependent 
Schr\"odinger equation. The argument of the function $f$ can be derived by 
considering the equation $i\partial \psi_{\vec k} (t) /\partial t = h_{\vec k}
\psi_{\vec k} (t)$, performing a phase re-definition to change $\Delta
({\vec k})$ to $|\Delta ({\vec k})|$, multiplying
both sides by $\sqrt{\tau/\epsilon ({\vec k})}$ and rescaling $t$ to
$t\sqrt{\tau/ \epsilon ({\vec k})}$. This effectively converts $h_{\vec k}$
to the form $~t\sigma^3_{\vec k} ~ +~ |\Delta({\vec k})| \sqrt{\tau/\epsilon
({\vec k})} (\sigma^+_{\vec k} + \sigma^-_{\vec k})$; hence the
probability $p_{\vec k}$ of starting in the ground state $\left(
\begin{array}{c}
1 \\ 0 \end{array} \right)$ at $t=-\infty$ and ending in the same state (which
is the excited state) at $t=\infty$ must be a function of $\tau |\Delta
({\vec k})|^2/\epsilon ({\vec k})$.

We can generalize the above results to a QCP with arbitrary
values of $z$ and $\nu$. We consider a generic time-dependent
Hamiltonian $H (t) \equiv H [\lambda(t)]$, whose states are labeled by
$|{\vec k} \rangle$, and $|0\rangle$ denotes the ground state. If there is a
second order phase transition, the basis states change continuously
with time during this evolution and can be written as $|\psi(t)\rangle =
\sum_{\vec k} a_{\vec k} (t) |{\vec k}[\lambda(t)]\rangle$. The defect
density can be obtained in terms of the coefficients $a_{\vec k} (t)$ as $n =
\sum_{{\vec k} \ne 0} |a_{\vec k} (t\to \infty)|^2$; hence one gets
\cite{mssanatoly1}
\begin{eqnarray} n ~\sim~ \int ~d^d k ~\Big| \int_{-\infty}^\infty d \lambda ~
\langle {\vec k}| \frac{d}{d \lambda} |0 \rangle ~e^{i \tau \int^\lambda d
\lambda' \delta \omega_{\vec k} (\lambda')}
\Big|^2, \label{mssdefect1} \end{eqnarray}
where $\delta \omega_{\vec k} (\lambda)=\omega_{\vec k} (\lambda)-
\omega_0(\lambda)$ are the
instantaneous excitation energies. Following Ref. \cite{mssanatoly1}, we note
that near a QCP, $\delta \omega_{\vec k} (\lambda) = \Delta F(\Delta /|
{\vec k}|^z)$, where $\Delta$ is the energy gap, $z$ is the dynamical 
critical exponent and $F(x) \sim 1/x$ for large $x$; we have assumed here 
that the gap vanishes as ${\vec k} = {\vec 0}$. Also, since the quench term 
vanishes at the critical point as $\Delta
\sim |\lambda|^{z \nu}$, one can write $\delta \omega_{\vec k} (\lambda)
= |\lambda|^{z \nu} {\tilde F} (|\lambda|^{z \nu}/|{\vec k}|^z)$ where
${\tilde F} (x) \sim 1/x$ for large $x$. Further, one has $\langle {\vec k}|
\frac{d}{d \Delta}|0\rangle = |{\vec k}|^{-z} G(\Delta
/|{\vec k}|^z)$ near a critical point where $G(0)$ is a constant. This allows
us to write $\langle {\vec k}|\frac{d}{d \lambda}|0\rangle =
\lambda^{z \nu -1} |{\vec k}|^{-z} G'(\lambda^{z \nu}/|{\vec k}|^z)$ where
$G'(0)$ is a constant \cite{msssubir1,mssanatoly1}.
Substituting these in (\ref{mssdefect1}) and changing
the integration variables to $\eta = \tau^{\nu/(z \nu + 1)} |{\vec k}|$
and $ \xi = |{\vec k}|^{-1/\nu} \lambda$, we find that
\begin{eqnarray} n ~\sim ~\tau^{-d \nu/(z \nu +1)}. \end{eqnarray}

We will now discuss two major extensions of the above results: (i) what
happens if the system is taken across a $d-m$ dimensional quantum critical
surface instead of a QCP \cite{mssmondal1}, and (ii) what happens if the
quenching across a QCP is non-linear in time \cite{mssmondal2,mssanatoly3}.
We will show that in both cases, the defect density still scales as a power
of the quench time, but the power is not equal to the universal value
$d\nu/(z \nu +1)$ mentioned above; rather it depends on other parameters
such as $m$ or the degree of non-linearity.

\subsection{Quenching across a critical surface}
\label{msscrit_surf}

When a quench takes a quantum system across a critical surface rather than a
critical point, the density of defects scales in a different way with the
quench time. To give a simple argument, consider a $d$-dimensional model 
with $z=\nu =1$ which is described by the Hamiltonian given in (\ref{msshfd}).
Suppose that a quench takes the system through a critical surface of $d-m$ 
dimensions. The defect density for a sufficiently slow quench is then given 
by \cite{msslz} $n \sim \int_{\rm BZ} d^d k e^{-\pi \tau f({\vec k})}$, where
$f ({\vec k})=|\Delta ({\vec k})|^2/|\epsilon ({\vec k})|$ vanishes on the 
$d-m$ dimensional critical surface. We can then write
\begin{equation} n ~\sim ~ \int_{\rm BZ} d^d k ~ \exp [~- \pi \tau
\sum_{\alpha, \beta=1}^m
g_{\alpha \beta} k_{\alpha} k_{\beta}] ~\sim~ 1/\tau^{m/2}, \end{equation}
where $\alpha, \beta$ denote one of the $m$ directions orthogonal to the
critical surface, and $g_{\alpha \beta} = [\partial^2 f({\vec k})/\partial
k_{\alpha} \partial k_{\beta}]_{{\vec k}
\in {\rm critical ~surface}}$. Note that this result depends only on the
property that $f({\vec k})$ vanishes on a $d-m$ dimensional surface, and not on
the precise form of $f({\vec k})$. For general values of $z$ and $\nu$, we note
that the Landau-Zener type of scaling argument yields $\Delta \sim
1/\tau^{d\nu/(z\nu+1)}$, where $\Delta$ is the energy gap \cite{mssanatoly1}.
When one crosses a $d-m$ dimensional critical surface during the quench,
the available phase space $\Omega$ for defect production scales as
$\Omega \sim k^m \sim \Delta^{m/z} \sim 1/\tau^{m \nu/(z\nu +1)}$;
this leads to $n \sim 1/\tau^{m \nu/(z\nu +1)}$. For a quench through a
critical point where $m=d$, we retrieve the results of Ref. \cite{mssanatoly1}.

\begin{figure}
\centerline{\rotatebox{0}{\includegraphics*[width=0.9\linewidth]{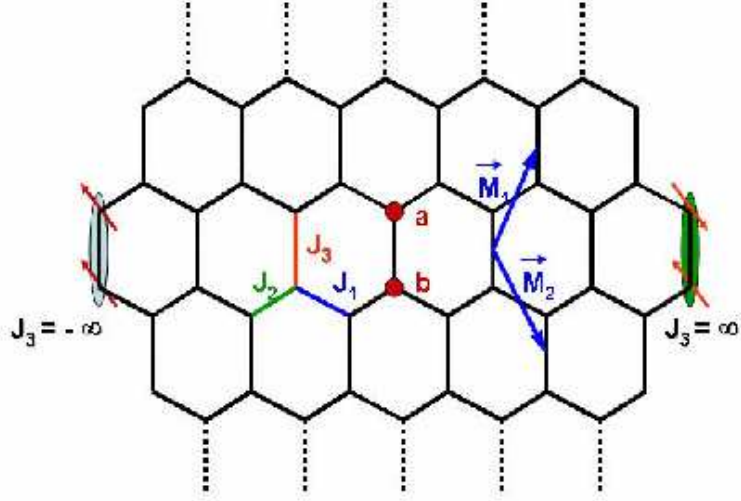}}}
\caption{Schematic representation of the Kitaev model on a honeycomb lattice
showing the bonds $J_1$, $J_2$ and $J_3$. Schematic pictures of the ground
states, which correspond to pairs of spins on vertical bonds locked parallel
(antiparallel) to each other in the limit of large negative (positive) $J_3$,
are shown at one bond on the left (right) edge respectively. ${\vec M}_1$ and
${\vec M}_2$ are spanning vectors of the lattice, and $a$ and $b$ represent
inequivalent sites.} \label{mssfig11} \end{figure}

To give an example of a quench across a critical line, let us consider a
model which was proposed recently by Kitaev. This is a 2D spin-1/2 model
on a honeycomb lattice as shown in Fig. \ref{mssfig11}; the Hamiltonian is
given by \cite{msskitaev1}
\begin{equation} H_K ~=~ \sum_{j+l={\rm even}} ~(~ J_1 \sigma_{j,l}^x
\sigma_{j+1,l}^x ~+~ J_2 \sigma_{j-1,l}^y \sigma_{j,l}^y ~+~ J_3
\sigma_{j,l}^z \sigma_{j,l+1}^z ~), \label{msskham1} \end{equation}
where $j$ and $l$ denote the column and row indices of the honeycomb lattice.
This model has been studied extensively and it exhibits several interesting
features \cite{mssfeng,mssbaskaran,msslee,mssvidal,mssnussinov1,mssnussinov2}.
It provides a rare example of a 2D model which can be exactly solved using a
Jordan-Wigner transformation
\cite{msskitaev1,mssfeng,mssnussinov1,mssnussinov2}. It has been shown
in Ref. \cite{msskitaev1} that the presence of magnetic field, which
induces a gap in the 2D gapless phase, leads to non-Abelian statistics of
the low-lying excitations of the model; these excitations can be viewed as
robust qubits in a quantum computer \cite{msskitaev2}.

The Jordan-Wigner transformation of the Kitaev model to a model of
non-interacting fermions works as follows. One can write
\begin{eqnarray} a_{jl} &=& \left( \prod_{i=-\infty}^{j-1} ~\sigma_{il}^z
\right) ~ \sigma_{jl}^y ~~{\rm for}~{\rm ~ even }~ j+l, \nonumber \\
b_{jl} &=& \left( \prod_{i=-\infty}^{j-1} ~\sigma_{il}^z \right)
~\sigma_{jl}^x ~~ {\rm for}~{\rm ~odd }~ j+l. \label{mssmaj2d} \end{eqnarray}
where the $a_{jl}$ and $b_{jl}$ are Majorana fermion operators (and hence
Hermitian) obeying the anticommutation relations $\{ a_{jl}, a_{j'l'} \} =
\{ b_{jl}, b_{j'l'} \} = \delta_{jj'} \delta_{ll'}$.
This transformation maps the spin Hamiltonian in (\ref{msskham1})
to a fermionic Hamiltonian given by
\begin{equation} H_K ~=~ i ~\sum_{\vec n} ~[J_1 ~b_{\vec n} a_{{\vec n}
- {\vec M}_1} ~+~ J_2 ~ b_{\vec n} a_{{\vec n} + {\vec M}_2} ~+~ J_3
D_{\vec n} ~b_{\vec n} a_{\vec n}], \label{msskham2} \end{equation}
where ${\vec n} = {\sqrt 3} {\hat i} ~n_1 + (\frac{\sqrt 3}{2} {\hat i} +
\frac{3}{2} {\hat j} ) ~n_2$ denote the midpoints of the vertical
bonds. Here $n_1, n_2$ run over all integers so that
the vectors ${\vec n}$ form a triangular lattice whose vertices
lie at the centers of the vertical bonds of the underlying honeycomb
lattice; the Majorana fermions $a_{\vec n}$ and $b_{\vec n}$ sit at the
top and bottom sites respectively of the bond labeled ${\vec n}$. The
vectors ${\vec M}_1 = \frac{\sqrt 3}{2} {\hat i} +
\frac{3}{2} {\hat j}$ and ${\vec M}_2 = \frac{\sqrt 3}{2} {\hat i} -
\frac{3}{2} {\hat j}$ are spanning vectors for the reciprocal lattice, and
$D_{\vec n}$ can take the values $\pm 1$ independently for each ${\vec n}$.
The crucial point that makes the solution of Kitaev model feasible is that
$D_{\vec n}$ commutes with $H_K$, so that all the eigenstates of $H_K$ can
be labeled by a specific set of values of $D_{\vec n}$. It has been
shown that for any value of the parameters $J_i$, the ground state of the
model always corresponds to $D_{\vec n}=1$ on all the bonds. Since
$D_{\vec n}$ is a constant of motion, the dynamics of the model starting
from any ground state never takes the system outside the manifold of states
with $D_{\vec n}=1$.

For $D_{\vec n}=1$, it is straightforward to diagonalize $H_K$ in momentum
space. We define Fourier transforms of the Majorana operators $a_{\vec n}$ as
\begin{eqnarray} a_{\vec n} &=& \sqrt{\frac{4}{N}} ~\sum_{\vec k} ~[~
a_{\vec k} ~e^{i{\vec k} \cdot {\vec n}} ~+~ a_{\vec k}^\dagger ~
e^{-i{\vec k} \cdot {\vec n}} ~], \nonumber \\
b_{\vec n} &=& \sqrt{\frac{4}{N}} ~\sum_{\vec k} ~[~ b_{\vec k} ~
e^{i{\vec k} \cdot {\vec n}} ~+~ b_{\vec k}^\dagger ~ e^{-i{\vec k} \cdot
{\vec n}} ~], \label{mssfourier2} \end{eqnarray}
where $N$ is the number of sites (assumed to be
even, so that the number of unit cells $N/2$ is an integer), and the sum over
${\vec k}$ extends over half the Brillouin zone of the honeycomb lattice. We
have the anticommutation relations $\{ a_{\vec k}, a_{{\vec k}'}^\dagger \}
= \delta_{{\vec k}, {\vec k}'}$, $\{ a_{\vec k}, a_{{\vec k}'} \} = 0$, and
similarly for $b_{\vec k}$ and $b_{\vec k}^\dagger$. We then obtain $H_K
= \sum_{\vec k} \psi_{\vec k}^\dagger h_{\vec k} \psi_{\vec k}$, where
$\psi_{\vec k}^\dagger =(a_{\vec k}^\dagger , b_{\vec k}^\dagger)$, and
$h_{\vec k}$ can be expressed in terms of Pauli matrices $\sigma^{1,2,3}$ as
\begin{eqnarray} h_{\vec k} &=& 2 ~[J_1 \sin ({\vec k} \cdot {\vec M}_1) ~
-~ J_2 \sin ({\vec k} \cdot {\vec M}_2)] ~\sigma^1 \nonumber \\
& & +~ 2 ~[J_3 ~+~ J_1 \cos ({\vec k} \cdot {\vec M}_1) ~+~ J_2 \cos
({\vec k} \cdot {\vec M}_2)] ~\sigma^2 . \end{eqnarray}
The energy spectrum of $H_K$ consists of two bands with energies
\begin{eqnarray} E_{\vec k}^\pm &=& \pm ~2 ~[(J_1 \sin ({\vec k} \cdot
{\vec M}_1) - J_2 \sin ({\vec k} \cdot {\vec M}_2))^2 \nonumber \\
& & ~~~~~~~~~+ (J_3 + J_1 \cos ({\vec k} \cdot {\vec M}_1) + J_2 \cos
({\vec k} \cdot {\vec M}_2))^2 ]^{1/2}. \end{eqnarray}
We note that for $|J_1-J_2|\le J_3 \le J_1+J_2$, these bands touch each
other so that the energy gap $\Delta_{\vec k} = E_{\vec k}^+ -
E_{\vec k}^-$ vanishes for special values of ${\vec k}$ leading to a
gapless phase of the model \cite{msskitaev1,mssfeng,msslee,mssnussinov1}.

We will now quench $J_3(t) =J t/\tau$ at a fixed rate $1/\tau$,
from $-\infty$ to $\infty$, keeping $J$, $J_1$ and $J_2$ fixed at some
non-zero values; we have introduced the quantity $J$ to fix the
scale of energy. We note that the ground states of $H_K$
corresponding to $J_3 \to -\infty(\infty)$ are gapped and
have $\sigma_{j,l}^z \sigma_{j,l+1}^z = 1(-1)$ for all lattice sites
$(j,l)$. To study the state of the system after the quench, we first
note that after an unitary transformation $U= \exp(-i \sigma_1 \pi/4)$, one
can write $H_K = \sum_{\vec k} \psi_{\vec k}^{'\dagger} h'_{\vec k}
\psi'_{\vec k}$, where $h'_{\vec k} = Uh_{\vec k} U^\dagger$ is given by
\begin{eqnarray} h'_{\vec k} &=& 2 ~[J_1 \sin ({\vec k} \cdot
{\vec M}_1) ~-~ J_2 \sin ({\vec k} \cdot {\vec M}_2)] ~\sigma^1 \nonumber \\
& & + ~2 ~[J_3(t) +J_1 \cos ({\vec k} \cdot {\vec M}_1) + J_2 \cos
({\vec k} \cdot {\vec M}_2)] ~\sigma^3 . \nonumber \\ \end{eqnarray}
Hence the off-diagonal elements of $h'_{\vec k}$ remain time-independent, 
and the problem of quench dynamics reduces to a
Landau-Zener problem for each ${\vec k}$. The defect density can then
be computed following a standard prescription \cite{msslz}
\begin{eqnarray} n &=& \frac{1}{A} ~\int_{\vec k} ~d^2 {\vec k} ~
p_{\vec k}, \nonumber \\
p_{\vec k} &=& e^{ - 2 \pi \tau ~[J_1 \sin ({\vec k} \cdot {\vec M}_1)- J_2
\sin ({\vec k} \cdot {\vec M}_2)]^2/J}, \label{mssdefect2d} \end{eqnarray}
where $A = 4\pi^2 /(3\sqrt{3})$ denotes the area of half the Brillouin zone
over which the integration is carried out. Since the integrand in
(\ref{mssdefect2d}) is an even function of ${\vec k}$, one can extend the
region of integration over the full Brillouin zone. This region can be chosen
to be a rhombus with vertices lying at $(k_x,k_y)= (\pm 2\pi /
\sqrt{3}, 0)$ and $(0,\pm 2\pi /3)$. Introducing two independent integration
variable $v_1, v_2$, each with a range $0\le v_1,v_2 \le 1$, one finds that
\begin{eqnarray} k_x &=& 2\pi ~\frac{v_1 + v_2 -1}{\sqrt 3}, \quad k_y = 2\pi ~
\frac{v_2 - v_1}{3}. \end{eqnarray}
Such a substitution covers the rhombus uniformly and facilitates the
numerical integration necessary for computing $n$.

\begin{figure}
\centerline{\rotatebox{0}{\includegraphics[width=3.4in]{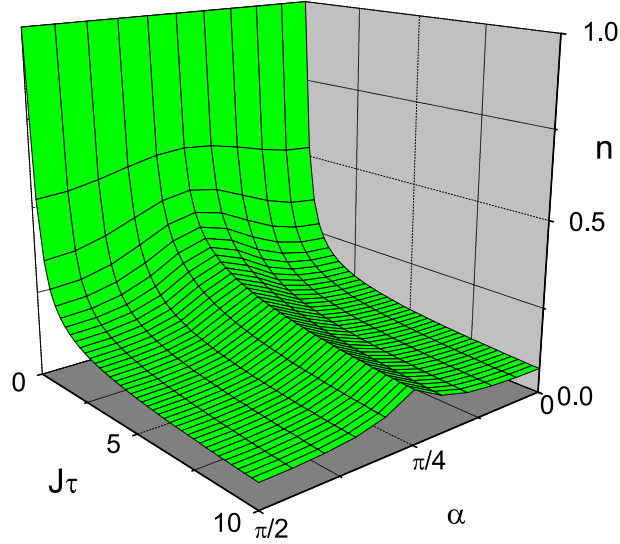}}}
\caption{Plot of defect density $n$ as a function of the quench time
$J \tau$ and $\alpha = \tan^{-1} (J_2/J_1)$. The density of defects
is maximum at $J_1=J_2$.} \label{mssfig12} \end{figure}

A plot of $n$ as a function of the quench time $J \tau$ and $\alpha =
\tan^{-1} (J_2/J_1)$ (we have taken $J_{1[2]}= J \cos \alpha [\sin \alpha ]$)
is shown in Fig. \ref{mssfig12}. We note
that the density of defects produced is maximum when $J_1=J_2$. This
is due to the fact that the length of the gapless line through which
the system passes during the quench is maximum at this point. This
allows the system to remain in the non-adiabatic state for the maximum time
during the quench, leading to the maximum density of defects. For $J_1/J_3
>2J_2/J_3$, the system does not pass through a gapless phase during the
quench, and the defect production is exponentially suppressed.

For sufficiently slow quench $2 \pi J \tau \gg 1$, $p_{\vec k}$ is
exponentially small for all values of ${\vec k}$ except in the region
near the line
\begin{equation} J_1 ~\sin ({\vec k} \cdot {\vec M}_1) ~-~ J_2 \sin
({\vec k} \cdot {\vec M}_2) ~=~ 0, \label{mssline} \end{equation} and the
contribution to the momentum integral in (\ref{mssdefect2d}) comes from
values of ${\vec k}$ close to this line of zeroes. We note that the line
of zeroes where $p_{\vec k}=1$ precisely corresponds to the zeroes of
the energy gap $\Delta_{\vec k}$ as $J_3$ is varied for a fixed $J_2$
and $J_1$. Thus the system becomes non-adiabatic when it passes through the
intermediate gapless phase in the interval $|J_1-J_2|\le J_3(t) \le
J_1+J_2$. It is then easy to see, by expanding $p_{\vec k}$ about
this line that in the limit of slow quench, the defect density scales as
$n \sim 1/\sqrt{\tau}$. We thus see that the scaling of the defect density
with the quench rate when the system passes through a critical {\it line} in
momentum space is different from the situation where the quench takes the
system through a critical {\it point}. The Kitaev model is an example of a
system in which $d=2$, $m=1$, and $z=\nu=1$; this gives rise to a defect
density scaling as $1/\sqrt{\tau}$.

\begin{figure}
\centerline{\rotatebox{0}{\includegraphics[width=3.4in]{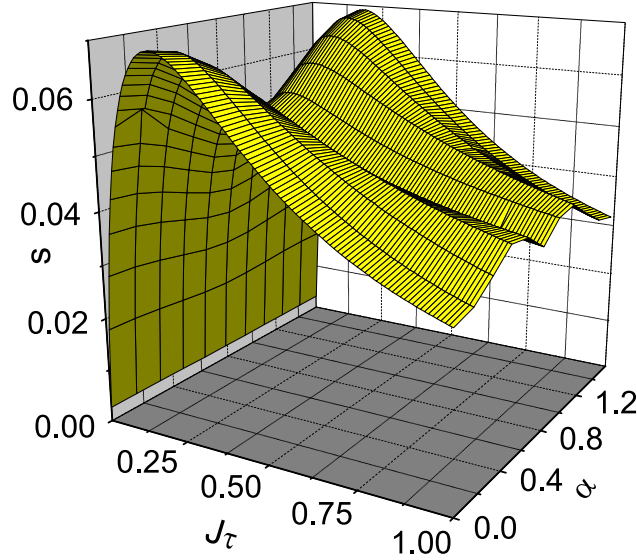}}}
\caption{Plot of the entropy density $s$ as a function of $J\tau$ and $\alpha=
\tan^{-1} (J_2/J_1)$. The entropy density peaks when $\left<O_{\vec 0}
\right>$ crosses from $-1$ to $1$.} \label{mssfig13} \end{figure}

Before ending this section, it is interesting to consider another aspect of
the system after the quench. Since the time evolution of the system is
unitary, it will always be in a pure state. However, for each value of $k$, 
the wave function is given by $\sqrt{1 - p_{\vec k}} \psi_{2{\vec k}} 
e^{-iE_{2k} t} ~+~ \sqrt{p_{\vec k}} \psi_{1{\vec k}}
e^{-iE_{1k} t}$, where $E_{1k}$ ($E_{2k}$) $= \infty$ ($-\infty$). As a
result, the final density matrix of the system will have off-diagonal terms
involving $\psi_{2{\vec k}}^* \psi_{1{\vec k}}$ and $\psi_{1{\vec k}}^*
\psi_{2{\vec k}}$ which vary extremely rapidly with time; their effects on
physical quantities will therefore average to zero. Hence, for each momentum
${\vec k}$, the final density
matrix $\rho_{\vec k}$ is effectively diagonal like that of a mixed state
\cite{msslevitov}, where the diagonal entries are time-independent as $t \to
\infty$ and are given by $1 - p_k$ and $p_k$. Such a density matrix is
associated with an entropy which we will now calculate. The density matrix of
the entire system takes the product form $\rho = \bigotimes \rho_{\vec k}$.
The von Neumann entropy density corresponding to this state is given by
\begin{equation} s ~=~ - ~\frac{1}{A} ~\int d^2 {\vec k} ~[~ (1 -
p_{\vec k}) \ln (1 - p_{\vec k}) ~
+~ p_{\vec k} \ln p_{\vec k} ~], \label{mssentropy1}\end{equation}
where the integral again goes half the Brillouin zone. Let us now consider
the dependence of this quantity on the quench time $\tau$
\cite{msssen1}. If $\tau$ is very small, the system stays in its initial
state and $p_{\vec k}$ will be close to 1 for all values of ${\vec k}$; for
the same reason, $\langle O_{\vec 0} \rangle$ will remain close
to 1. If $\tau$ is very large, the system makes a transition to the
final ground state for all momentum except near the line described in
(\ref{mssline}). Hence $p_{\vec k}$ will be close to 0 for all ${\vec k}$
except near that line, and $\langle O_{\vec 0} \rangle$ will be close
to -1. In both these cases, the entropy density will be small. We
therefore expect that there will be an intermediate region of values
of $\tau$ in which $s$ will show a maximum and $\langle O_{\vec 0} \rangle$
will show a crossover from $-1$ to 1. A plot of $s$ as a
function of $J\tau$ and $\alpha$ shown in Fig. \ref{mssfig13} confirms
this expectation. We find that the entropy reaches a maximum for an
intermediate value of $J\tau$ where $\langle O_{\vec 0} \rangle$ crosses
over from $-1$ to 1 for all values of $\alpha$.

\subsection{Non-linear quenching across a critical point}
\label{mssnon-linear}

Let us now consider what happens if we start with a Hamiltonian similar to
the one given in (\ref{msshfd}), except that
\begin{equation} h_{\vec k} (t) ~=~ (\lambda(t)+b({\vec k}))
\sigma^3_{\vec k} ~+~ \Delta({\vec k}) ~\sigma^+_{\vec k} ~+~
\Delta^* ({\vec k}) ~\sigma^-_{\vec k} , \label{msshfd2} \end{equation}
where $\lambda(t)= \lambda_0 |t/\tau|^{\alpha} {\rm sign}(t)$ is the quench
parameter; $\alpha=1$ corresponds to a linear quench. The instantaneous
energies of the Hamiltonian in (\ref{msshfd2}) are given by
\begin{eqnarray} E({\vec k}) &=& \pm ~\sqrt{(\lambda(t)+b({\vec k}))^2 + |
\Delta ({\vec k})|^2}. \label{mssen1} \end{eqnarray}
These energy levels touch each other at $t=t_0$ and ${\vec k}={\vec k}_0$, so
that $|\Delta ({\vec k})| \sim |{\vec k} - {\vec k}_0|$ and $|t_0| = \tau
|b({\vec k}_0)/\lambda_0|^{1/\alpha} = \tau g^{1/\alpha}$, where $g=
|b({\vec k}_0)/\lambda_0|$ is a non-universal model-dependent parameter. At
this point the energy levels cross and we have a QCP with $z=\nu=1$. Note
that the critical point is reached at $t=0$ only if $b({\vec k}_0)$ vanishes.

Let us first consider the case where $b({\vec k}_0)=0$ so that the system
passes through the critical point at $t=0$. In what follows, we shall assume
that $|\Delta ({\vec k})| \sim |{\vec k} -{\vec k}_0|$ and $ b({\vec k})
\sim |{\vec k}-{\vec k}_0|^{z_1}$ at the critical point, where $z_1 \ge 1$ so
that $E \sim |{\vec k}-{\vec k}_0|$ and $z=1$. In the rest of the analysis, we
will scale $t \to t\lambda_0$, $\tau \to \tau \lambda_0$, $\Delta ({\vec k})
\to \Delta ({\vec k})/\lambda_0$, and $b({\vec k}) \to b({\vec k})/\lambda_0$.

To obtain the probability $p_{\vec k}$ of ending in the excited state at
$t = \infty$, we study the time evolution of the system
governed by the Schr\"odinger equation $i\partial \psi_{\vec k}
/\partial t =h_{\vec k} \psi_{\vec k}$. This leads to the equations
\begin{eqnarray} i\dot{c}_{1{\vec k}} &=& (|t/\tau|^{\alpha} {\rm sign}(t)
+b({\vec k})) ~c_{1{\vec k}} ~+~ \Delta ({\vec k}) ~c_{2{\vec k}}, \nonumber \\
i\dot{c}_{2 {\vec k}} &=& -(|t/\tau|^{\alpha} {\rm sign}(t) +b({\vec k})) ~
c_{2{\vec k}} ~+~ \Delta^* ({\vec k}) ~c_{1{\vec k}}, \label{mssc2}
\end{eqnarray}
where $\dot{c}_{1{\vec k}(2 {\vec k})} \equiv \partial_t c_{1{\vec k}
(2{\vec k})}$. To solve these equations, we define
\begin{eqnarray} c_{1{\vec k}}' &=& c_{1{\vec k}} ~e^{i\int^t dt' (| t'/
\tau|^\alpha {\rm sign} (t') + b({\vec k}))}, \nonumber \\
c_{2{\vec k}}' &=& c_{2 {\vec k}} ~e^{-i\int^t dt'(|t'/\tau|^\alpha
{\rm sign} (t') + b({\vec k}))}. \label{mssc2'} \end{eqnarray}
Substituting (\ref{mssc2'}) in (\ref{mssc2}) and eliminating
$c_{2{\vec k}}'$ from the resulting equations, we get
\begin{equation} \ddot{c}_{1{\vec k}}' ~-~ 2i ~[|t/\tau|^{\alpha}{\rm sign}
(t)+b({\vec k})] ~\dot{c}_{1{\vec k}}' ~+~ |\Delta ({\vec k})|^2 ~c_{1
{\vec k}}' ~=~ 0. \label{mssddotc1} \end{equation}
Now we scale $t \to t\tau^{\alpha/(\alpha+1)}$ so that (\ref{mssddotc1})
becomes
\begin{equation} \ddot{c}_{1{\vec k}}' ~-~ 2i ~[|t|^\alpha {\rm sign}(t) +
b({\vec k})\tau^{\alpha/ (\alpha +1)}] ~\dot{c}_{1{\vec k}}' ~+~ |\Delta
({\vec k})|^2 \tau^{2\alpha/(\alpha +1)} ~
c_{1{\vec k}}' ~=~ 0. \label{mssddotc11} \end{equation}
{}From (\ref{mssddotc11}) we note that since $c_{1{\vec k}}$ and
$c_{1{\vec k}}'$ differ only by a phase factor, $p_{\vec k}$ must be
of the form
\begin{equation} p_{\vec k} ~=~ f[b({\vec k}) \tau^{\frac{\alpha}{\alpha
+1}},|\Delta ({\vec k})|^2 \tau^{\frac{2\alpha}{\alpha+1}}], \label{msspexp}
\end{equation}
where $f$ is a function whose analytical form is not known for $\alpha \ne
1$. Nevertheless, we note that for a slow quench (large $\tau$),
$p_{\vec k}$ becomes appreciable only when the instantaneous energy gap,
as obtained from (\ref{mssen1}), becomes small at some point of time during
the quench. Consequently, $f$ must vanish when either of its arguments are
large: $f(\infty,a)=f(a,\infty)=0$ for any value of $a$. Thus for a
slow quench (large $\tau$), the defect density $n$ is given by
\begin{equation} n ~\sim~ \int_{\rm BZ} ~d^d k ~f [b({\vec k})
\tau^{\frac{\alpha}{\alpha+1}},|\Delta ({\vec k})|^2\tau^{\frac{2\alpha}{
\alpha+1}} ], \end{equation}
and receives its main contribution from values of $f$ near ${\vec k}={\vec
k}_0$ where both $b({\vec k})$ and $\Delta ({\vec k})$ vanish. Thus one
obtains, after extending the range of the integration to $\infty$,
\begin{equation} n ~\sim ~\int ~d^d k ~f\left[|{\vec k}-{\vec k}_0|^{z_1}
\tau^{\frac{\alpha}{\alpha+1}}; |{\vec k}-{\vec k}_0|^2
\tau^{\frac{2\alpha}{\alpha+1}}\right]. \end{equation}
Now scaling ${\vec k} \to ({\vec k} - {\vec k}_0) \tau^{\alpha/(\alpha+1)}$,
we find that
\begin{eqnarray} n &\sim& \tau^{- \frac{d \alpha}{\alpha+1}} ~\int ~d^d k~
f(|{\vec k}|^{z_1} \tau^{\alpha(1-z_1)/(\alpha+1)};|{\vec k}|^2) \nonumber \\
&\sim& \tau^{- \frac{d \alpha}{\alpha+1}} ~\int ~d^d k~ f(0;|{\vec k}|^2)~
\sim ~\tau^{- \frac{d \alpha}{\alpha+1}}, \label{msssca1} \end{eqnarray}
where, in arriving at the last line, we have used $z_1 > 1$ and $\tau
\to \infty$. (If $z_1 = 1$, the integral in the first line is
independent of $\tau$, so the scaling argument still holds). Note
that for $\alpha=1$, Eq. (\ref{msssca1}) reduces to its counterpart for a
linear quench \cite{mssanatoly1}. It turns out that the case $z_1 < 1$
deserves a detailed discussion which is given in Ref. \cite{mssmondal2}.

Next we generalize our results to a critical point with arbitrary values of
$z$ and $\nu$. We use arguments similar to those given in the discussion
around Eq. (\ref{mssdefect1}), namely,
\begin{eqnarray} n ~\sim~ \int ~d^d k ~\Big| \int_{-\infty}^\infty d \lambda ~
\langle {\vec k}| \frac{d}{d \lambda} |0 \rangle ~e^{i \tau \int^\lambda d
\lambda' \delta \omega_{\vec k} (\lambda')}
\Big|^2. \label{mssdefect2} \end{eqnarray}
In the present case, the quench term vanishes at the critical point as
$\Delta \sim |\lambda|^{\alpha z \nu}$ for a nonlinear quench, and we can write
\begin{equation} \delta \omega_{\vec k} (\lambda) ~=~ |\lambda|^{\alpha z
\nu} {\tilde F} (|\lambda|^{\alpha
z\nu}/ |{\vec k}- {\vec k}_0|^z), \label{mssrel2} \end{equation}
where ${\tilde F} (x) \sim 1/x$ for large $x$. Further, $\langle {\vec k}|
\frac{d}{d \Delta} |0\rangle = |{\vec k}-{\vec k}_0|^{-z} G(\Delta /|
{\vec k}-{\vec k}_0|^z)$ near a
critical point, where $G(0)$ is a constant. This allows us to write
\begin{eqnarray} \langle {\vec k}|\frac{d}{d \lambda}|0\rangle ~=~ \frac{
\lambda^{\alpha z \nu -1}}{|{\vec k}- {\vec k}_0|^z} G'(\lambda^{\alpha z
\nu}/|{\vec k}-{\vec k}_0|^z), \label{mssrel3} \end{eqnarray}
where $G'(0)$ is a constant \cite{msssubir1,mssanatoly1}. Substituting
(\ref{mssrel2}) and (\ref{mssrel3})
in (\ref{mssdefect2}) and changing the integration variables to $\eta =
\tau^{\alpha \nu/(\alpha z \nu + 1)} |{\vec k}-{\vec k}_0|$ and
$ \xi = |{\vec k}-{\vec k}_0|^{-1/(\alpha \nu)} \lambda$, we find that
\begin{equation} n ~\sim~ \tau^{-\alpha d \nu/(\alpha z \nu +1)}.
\label{mssdefect3} \end{equation}

Next we consider the case where the quench term does not vanish at
the QCP for ${\vec k} = {\vec k}_0$. We again consider the
Hamiltonian $h_{\vec k} (t)$ in (\ref{msshfd2}), but now assume that the
critical point is reached at $t=t_0 \ne 0$. This renders our
previous scaling argument invalid since $\Delta ({\vec k}_0) = 0$ but
$b({\vec k}_0) \ne 0$. In this situation, $|t_0/\tau| = g^{1/\alpha}$ so
that the energy gap $\Delta E$ may vanish at the critical point for
${\vec k} = {\vec k}_0$. We now note that the most important contribution to
the defect production comes from times near $t_0$ and from wave numbers
near $k_0$. Hence we expand the diagonal terms in $h_{\vec k} (t)$
about $t=t_0$ and ${\vec k} ={\vec k}_0$ to obtain
\begin{equation} H ~=~ \sum_{\vec k} ~\Big[ \left\{ \alpha g^{(\alpha-1)/
\alpha} \left( \frac{t-t_0}{\tau}\right)+ b' (\delta {\vec k}) \right\}
\sigma_{\vec k}^3 ~+~ \Delta({\vec k}) \sigma_{\vec k}^+ ~+~ \Delta^*
({\vec k}) \sigma_{\vec k}^- \Big], \label{mssham3} \end{equation}
where $b'(\delta {\vec k})$ represents all the terms in the expansion of
$b({\vec k})$ about ${\vec k}={\vec k}_0$, and we have neglected all terms
\begin{equation} R_n ~=~ (\alpha-n+1)(\alpha-n+2)...(\alpha) ~g^{(\alpha-n)/
\alpha} |(t-t_0)/\tau|^n {\rm sign}(t)/n! \label{mssnegterm1} \end{equation}
for $n>1$ in the expansion of $\lambda (t)$ about $t_0$. We shall justify
neglecting these higher order terms shortly.

Eq. (\ref{mssham3}) describes a linear quench of the system with
$\tau_{\rm eff}(\alpha) = \tau/(\alpha g^{(\alpha-1)/\alpha})$. Hence one
can use the well-known results of Landau-Zener dynamics \cite{msslz} to
write an expression for the defect density,
\begin{equation} n ~\sim~ \int_{\rm BZ} ~d^d k ~p_{\vec k} ~\sim~
\int_{\rm BZ}~ d^d k ~ \exp[ -\pi |\Delta ({\vec k})|^2 \tau_{\rm eff}
(\alpha)]. \label{mssefq1} \end{equation}
For a slow quench, the contribution to $n$ comes from ${\vec k}$ near
${\vec k}_0$; hence
\begin{equation} n ~\sim ~\tau_{\rm eff}(\alpha)^{-d/2} ~=~ \left( \alpha
g^{(\alpha-1)/\alpha}/\tau \right)^{d/2}. \label{mssdefect4} \end{equation}
Note that for the special case $\alpha=1$, we recover the familiar result $n
\sim \tau^{-d/2}$, and the dependence of $n$ on the non-universal constant
$g$ vanishes. Also, since the quench is effectively linear, we can use the
results of Ref. \cite{mssanatoly1} to find the scaling of the defect
density when the critical point at $t=t_0$ is characterized by arbitrary
$z$ and $\nu$,
\begin{equation} n ~\sim ~\left(\alpha g^{(\alpha-1)/\alpha}/\tau\right)^{\nu
d/(z\nu +1)}. \label{mssdefect5} \end{equation}

Next we justify neglecting the higher order terms $R_n$. We note that
significant contributions to $n$ come at times $t$ when the instantaneous
energy levels of $H$ in (\ref{mssham3}) for a given ${\vec k}$
are close to each other, i.e., $(t-t_0)/\tau \sim \Delta ({\vec k})$. Also,
for a slow quench, the contribution to the defect density is substantial
only when $p_{\vec k}$ is significant, namely, when $|\Delta ({\vec k})|^2
\sim 1/\tau_{\rm eff}(\alpha)$. Using these arguments, we see that
\begin{equation} R_n/R_{n-1} ~=~ (\alpha-n+1)g^{-1/\alpha}(t-t_0)/(n\tau) ~
\sim~ (\alpha-n+1)/(n \sqrt{\tau}). \label{mssnegterm2} \end{equation}
Thus we find that all higher order terms $R_{n>1}$, which were neglected in
arriving at (\ref{mssdefect4}), are unimportant in the limit of slow quench
(large $\tau$).

The scaling relations for the defect density $n$ given by
(\ref{mssdefect3}) and (\ref{mssdefect5}) represent the central results of
this section. For such power law quenches, unlike their linear
counterpart, $n$ depends crucially on whether or not the quench term
vanishes at the critical point. For quenches which do not vanish at
the critical point, $n$ scales with the same exponent as that of a
linear quench, but is characterized by a modified non-universal
effective rate $\tau_{\rm eff}(\alpha)$. If, however, the quench term
vanishes at the critical point, we find that $n$ scales with a novel
$\alpha$-dependent exponent $ \alpha d \nu/(\alpha z \nu +1)$. For $\alpha=1$,
$\tau_{\rm eff}(\alpha) = \tau$ and $\alpha d \nu/(\alpha z \nu + 1)
= d \nu/(z\nu +1)$; hence both (\ref{mssdefect3}) and
(\ref{mssdefect5}) reproduce the well-known defect production law for
linear quenches as a special case \cite{mssanatoly1}. We note that the
scaling of $n$ will show a cross-over between the expressions given
in (\ref{mssdefect3}) and (\ref{mssdefect5}) near some value of $\tau
= \tau_0$ which can be found by equating these two expressions; this
yields $\tau_0 \sim |b ({\vec k}_0)|^{- z \nu - 1/\alpha}$. For $\alpha
> 1$, the scaling law will thus be given by Eq. (\ref{mssdefect3})
((\ref{mssdefect5})) for $\tau \ll(\gg) \tau_0$. We also note here
that the results of this section assumes that the system passes from
one gapped phase to another through a critical point and do not
apply to quenches which take a system along a critical line
\cite{mssmondal1,msspell1,mssdiva1,mssdiva2}.

\begin{figure}
\centerline{\rotatebox{0}{\includegraphics[width=3.4in]{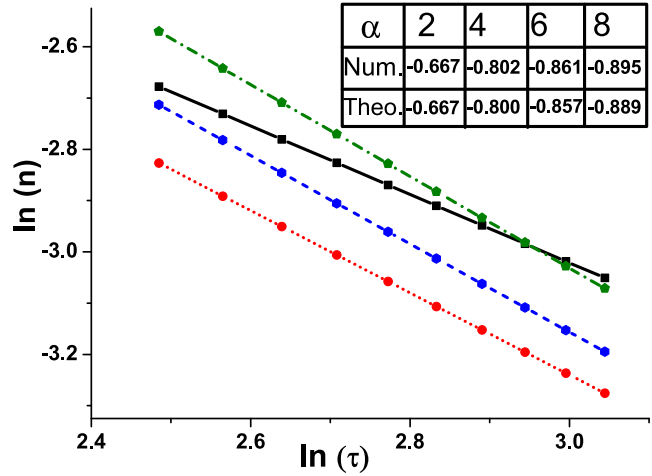}}}
\caption{(Color online) Plot of $\ln(n)$ vs $\ln(\tau)$ for the 1D
Kitaev model for $\alpha=2$ (black solid line), $\alpha=4$ (red dotted line),
$\alpha=6$ (blue dashed line) and $\alpha=8$ (green dash-dotted line). The
slopes of these lines agree reasonably with the predicted theoretical values
$-\alpha/(\alpha+1)$ as shown in the table.} \label{mssfig14} \end{figure}

To illustrate the form of defect scaling for a non-linear quench, let us
consider the 1D spin-1/2 Kitaev model which is governed by the Hamiltonian
\begin{equation} H ~=~ \sum_{i\in \rm{even}} ~\left( J_1 S_i^x S_{i+1}^x ~+~
J_2 S_i^y S_{i-1}^y \right), \end{equation}
where $S_i^a = \sigma_i^a /2$. Using the standard Jordan-Wigner transformation,
this can be mapped to a Hamiltonian of non-interacting fermions
\begin{eqnarray} H &=& \sum_{\vec k} ~\psi^{\dagger}_k ~h_k ~\psi_k,
\nonumber \\
{\rm where} ~~~h_k &=& -2~(J_-\sin k ~\tau_3+J_+\cos k ~\tau_2). \end{eqnarray}
Here $J_{\pm}=J_1\pm J_2$, and $\psi_k=(c_1(k),c_2(k))$ are the fermionic
fields. We now perform a quench by keeping $J_+$ fixed and varying the
parameter $J_-$ with time as $J_-(t)= J |t/\tau|^\alpha {\rm sign}(t)$. We
then pass through a QCP at $t=0$ at the wave number $k=\pi/2$. From
Eq. (\ref{mssdefect3}) we expect the defect density to go as
$n\sim\tau^{-\alpha/(\alpha+1)}$ since $\nu=z=1$ for this
system. To check this prediction, we numerically solve the
Schr\"odinger equation $i \partial \psi(k,t) /\partial t = h_k (t) \psi(k,t)$
and compute the defect density $n ~=~ \int_0^{\pi} ~(dk/\pi) ~p_k$ as
a function of the quench rate $\tau$ for different $\alpha$, with fixed $J_+
/J=1$. A plot of $\ln (n)$ vs $\ln (\tau)$ for
different values of $\alpha$ is shown in Fig. \ref{mssfig14}. The slopes of
these lines, as can be seen from Fig. \ref{mssfig14}, changes from
$-0.67$ towards $-1$ as $\alpha$ increases from $2$ towards larger values.
This behavior is consistent with the prediction of (\ref{mssdefect3}).
The slopes of these lines show excellent agreement with
(\ref{mssdefect3}) as shown in the inset of Fig. \ref{mssfig14}.

To illustrate what happens if the QCP is crossed at a time $t$ which is
different from 0, we consider the 1D Ising model in a transverse magnetic
field described by
\begin{equation} H_{\rm Ising} ~=~ - ~J ~(\sum_i ~S_i^z S_{i+1}^z ~+~ g ~
\sum_i ~S_i^x), \end{equation}
where $J$ is the strength of the nearest neighbor interaction, and $g=h/J$ is
the dimensionless transverse field. In what follows, we shall quench the
transverse field as $g(t)= |t/\tau|^{\alpha} {\rm sign}(t)$ and compute the
density of the resultant defects.

\begin{figure}
\centerline{\rotatebox{0}{\includegraphics*[width=0.7\linewidth]{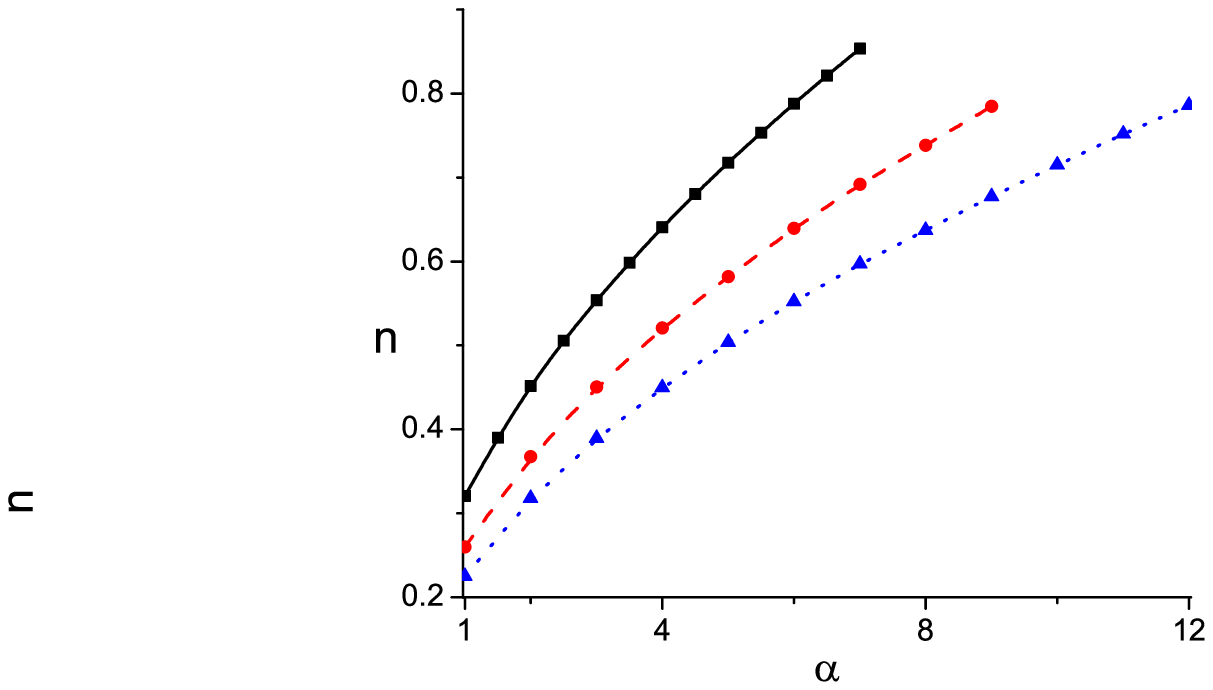}}}
\caption{(Color online) Variation of the defect density $n$ with the quench
exponent $\alpha$ for representative values of $\tau=10$ (black solid line),
$\tau=15$ (red dashed line) and $\tau=20$ (blue dotted line). A polynomial
fit of the form $n = a \alpha^b$ yields exponents which are very close to the
theoretical result $1/2$ for all values of $\tau$.} \label{mssfig15}
\end{figure}

We begin by mapping $H_{\rm Ising}$ to a system of free fermions via
the Jordan-Wigner transformation
\begin{equation} H ~=~ - ~J ~\sum_k ~[(g ~-~ \cos k) ~\sigma_k^3 ~+~ \sin k ~
\sigma_k^1]. \end{equation}
If the field $g$ is varied with time as $g(t)=g_0 |t/\tau|^{\alpha}
{\rm sign}(t)$, the system will go through two QCPs at $g= 1$ and $-1$.
The energy gap vanishes at these QCPs at $k=k_0=0$ and
$\pi$. As a result, defects are produced in non-adiabatic regions near these
points. For this model, the QCP is at $t=t_0\ne 0$ and $z= \nu=1$. Hence,
$\tau_{\rm eff} = \tau/\alpha$ for both the QCPs. From
(\ref{mssdefect5}), therefore, we expect the defect density
produced in this system to be given by $n\sim(\tau/\alpha)^{-1/2}$.

To verify this, we numerically solve the Schr\"odinger
equation $i\partial \psi_k (t) /\partial t = h_k (t) \psi_k (t)$ and obtain
the probability $p_k$ for the system to be in the excited state.
Finally, integrating over all $k$ within the Brillouin zone, we
obtain the defect density $n$ for different values of $\alpha> 1$ with
fixed $\tau$. The plot of $n$ as a function of $\alpha$ for
$\tau=10,~15$ and $20$ is shown in Fig. \ref{mssfig15}. A fit to these
curves gives the values of the exponents to be $0.506 \pm 0.006$,
$0.504 \pm 0.004$ and $0.505 \pm 0.002$ for $\tau=10,~15$ and $20$
respectively which are remarkably close to the theoretical value
$1/2$. The systematic positive deviation of the exponents from the
theoretical value $1/2$ comes from the contribution of the higher
order terms neglected in the derivation of (\ref{mssdefect4}) and
(\ref{mssdefect5}). We note that the region of validity of our linear
expansion, as can be seen from Fig. \ref{mssfig15}, grows with $\tau$
which is in accordance with the result in (\ref{mssnegterm2}).

\subsection{Experimental realizations}
\label{mssexpt}

The validity of our results can be checked in a variety of experimental
systems. We first observe that all our results
have been obtained at zero temperature with the assumption that the
system does not relax significantly during the quenching process and until
the defect density has been measured. This might seem rather restrictive.
We note however that systems of ultracold atoms in optical or magnetic traps
and/or optical lattices can easily satisfy the required criteria since they
have a very long relaxation time which often gets close to the system
lifetime \cite{mssbloch2}. We will briefly list some possible experiments
here. First, there has been a proposal for realizing
the Kitaev model using an optical lattice\cite{mssduan}. In such a
realization, all the couplings can be independently tuned using
separate microwave radiations. In the proposed experiment, one needs
to keep $J_3=0$ and vary $J_{1(2)} = J(1\pm |t/\tau|^{\alpha} {\rm
sign}(t))/2$, so that $J_+$ remains constant while $J_-$ varies in
time. The variation of the defect density, which in the experimental
set-up would correspond to the bosons being in the wrong spin state,
would then show the theoretically predicted power law behavior in
(\ref{mssdefect3}). Secondly, a similar quench experiment can be carried out
with spin-1 bosons in a magnetic field described by an effective Hamiltonian
$H_{\rm eff} = c_2 n_0 \langle {\bf S} \rangle^2 + c_1 B^2 \langle S_z^2
\rangle~$ \cite{msssadler}, where $c_2 < 0$ and $n_0$ is the boson
density. Such a system undergoes a quantum phase transition from a
ferromagnetic state to a polar condensate at $B^* =
\sqrt{|c_2|n_0/c_1}$. A quench of the magnetic field $B^2=B_0^2
|t/\tau|^{\alpha}$ would thus lead to a scaling of the defect density with an
effective rate $\tau_{\rm eff} (\alpha)= \tau/(\alpha g^{(\alpha-1)/\alpha})$,
where $g=|c_2|n_0 /c_1$. A measurement of the dependence of the
defect density $n$ on $\alpha$ should therefore serve as a test of the
prediction in (\ref{mssdefect5}). Finally, spin gap dimer compounds such
as ${\rm Ba Cu Si_2 O_6}$ are known to undergo a singlet-triplet quantum phase
transition of the Bose-Einstein condensation type at $B_c \simeq 23 $T; the
critical exponents for this are given by $z=2$ and $\nu = 2/d$. Experimentally,
the exponent $\nu$ appears to be $2/3$ above a temperature window of $0.65$K
to $0.9$K \cite{msssebas1} and $1$ below that temperature window due to a
dimensional reduction from $d=3$ to $d=2$ \cite{msssebas2}. Thus a nonlinear
quench of the magnetic field through its critical value $B=B_c + B_0
|t/\tau|^{\alpha}{\rm sign}(t)$ should lead to a scaling of the defects $n
\sim \tau^{-6\alpha/(4\alpha+3)}$ for $d=3$, $\nu = 2/3$, and $n \sim
\tau^{-2\alpha/(2\alpha+1)}$ for $d=2$, $\nu = 1$. It would be interesting to
see if the defect scaling exponent depends on the temperature range in the
same way as the exponent $\nu$. In the experiment, the defect density would
correspond to residual singlets in the final state which can be computed by
measuring the total magnetization of the system immediately after the quench. 
We note that for these dimer systems, it would be necessary to take special 
care to achieve the criterion of long relaxation time mentioned earlier.

\section{Quantum communication}
\label{mssquantum:comm}

In this section, we demonstrate that a properly engineered
non-adiabatic dynamics may lead to larger fidelity and higher speed
for the transfer of a qubit through a system. For this purpose, we begin with
a Heisenberg spin-1/2 chain described by a generic time-dependent Hamiltonian
\begin{equation} H ~=~ -J_0(t) ~\sum_{ij} (S_i^x S_j^x + S_i^y S_j^y) ~+~
\Delta(t) \sum_{ij} S_i^z S_j^z ~+~ B(t) ~\sum_i S_i^z . \label{mssham5} 
\end{equation}
We assume that the spin system is on a ring with $N$ sites.
We start with the initial ground state being ferromagnetic, and
denote this state by $|G\rangle$. At the start of the procedure of
qubit transfer, we put a state $ \cos(\theta/2) |\uparrow\rangle +
\sin(\theta/2) \exp(i \phi) |\downarrow\rangle$ at the $r^{\rm th}$
site of the chain. Thus the initial state of the system at the start
of the evolution is \cite{mssbose1}
\begin{eqnarray} |\psi_{\rm in}\rangle &=& \cos(\theta/2) |G\rangle +
\sin(\theta/2) e^{i\phi} |r\rangle, \end{eqnarray}
where $|r\rangle$ denotes the state of the spin chain with one
flipped spin at the site $r$. We now consider the evolution of this
state under a time-dependent Hamiltonian $H$. The specific form of
the interaction need not be specified at the moment. Since the total
spin is a conserved quantity ($\left[ \sum_i S_i^z, H\right]=0$),
the state of the system at time $t$ becomes
\begin{eqnarray} |\psi(t)\rangle &=& \cos(\theta/2) |G\rangle + \sin(\theta/2)
e^{i\phi} \sum_n f_{n r}(t) |n\rangle , \nonumber \\
{\rm where} ~~~~f_{n r}(t) &=& \langle n | e^{- i \int^t H(t') dt'} |r\rangle .
\end{eqnarray}
Since the idea of communication through the chain involves
performing measurement on the state at site $s^{\rm th}$ site, we
would like to compute the reduced density matrix of this site at
time $t$. To this end, we write the wave function
\begin{equation} |\psi(t)\rangle = \cos(\theta/2) |G\rangle + \sin(\theta/2)
e^{i\phi} \sum_{n\ne s} f_{n r}(t) |n\rangle + \sin(\theta/2) e^{i\phi}
f_{sr}(t) |s\rangle , \label{msswave1} \end{equation}
where the first line of the last equation is the contribution from
all terms to $|\psi(t)\rangle$ where the spin in the $s^{\rm th}$ site is
$\uparrow$. Note that for normalization of the wave function, one needs
\begin{equation} \cos^2(\theta/2) + \sin^2(\theta/2) \sum_{n\ne s} |f_{n r}
(t)|^2 = 1- |f_{sr}(t)|^2 \sin^2(\theta/2). \label{mssnorm} \end{equation}
Using (\ref{msswave1}) and (\ref{mssnorm}), one find that the reduced
density matrix for the $s^{\rm th}$ site of the system is
\begin{eqnarray} \rho_s(t) &=& (1-|f_{sr}(t)|^2 \sin^2(\theta/2)) ~|\uparrow
\rangle \langle \uparrow | ~+~ |f_{sr}(t)|^2 \sin^2(\theta/2) ~|\downarrow
\rangle \langle \downarrow | \nonumber \\
& & + ~\frac{\sin(\theta)}{2} ~\left( e^{i \phi} f_{sr}(t) |\downarrow\rangle
\langle \uparrow | ~+~ e^{-i \phi} f_{sr}^* (t) |\uparrow\rangle \langle
\downarrow | \right). \end{eqnarray}
The fidelity of the state transfer at the given time $t$ is thus defined as 
\cite{mssbose1}
\begin{eqnarray} F(t) &=& \frac{1}{4\pi} \int d\Omega \langle \psi_{\rm in} |
\rho_s(t) | \psi_{\rm in} \rangle \nonumber \\
&=& \frac{1}{2} + \frac{|f_{sr}(t)|^2}{6} + \frac{Re[f_{rs}(t)]}{3},
\label{mssfeq} \end{eqnarray}
where the integration is over the Bloch sphere involving $\theta$
and $\phi$. Thus to obtain the fidelity of a state transfer we need
to obtain the matrix elements $f_{sr}(t)$. To do this, we note that
since the Hamiltonian in (\ref{mssham5}) conserves the $z$ component of
the spin, an arbitrary time-dependent dynamics always restricts the system
to lie within the subspace of one flipped spin. This allows us to write the
wave function after an evolution through a time $t$ to be
\begin{eqnarray} |\phi(t)\rangle &=& \sum_n c_n(t) |n\rangle = \sum_k
c_k (t) |k\rangle, \label{msswave2} \end{eqnarray}
where the real space basis $|n\rangle$ and the wave number space basis
$|k\rangle$ are related by $|n\rangle = \sum_k \exp(-i k n)
|k\rangle$ for a chain with a periodic boundary condition. The
Schr\"odinger equation for $|\phi(t)\rangle$ now leads to the
following equation for $c_k(t)$
\begin{eqnarray} i \dot c_k(t) &=& \left(2 J(t) \cos(k) + \frac{1}{4}
[\Delta(t)+2 B(t)] \right) c_k(t), \label{mssceq1} \end{eqnarray}
where we have neglected factors of $1/N$ ($N$ being the chain
length which approaches infinity in the thermodynamic limit) in the
expression for $\beta (t)$. These equations are to be solved with the
boundary condition $c_n(t=0) = \delta_{nr}$. This equation has a
straightforward solution
\begin{eqnarray} c_k(t) &=& e^{-i(2 \alpha (t) \cos(k) + \beta (t))},
\nonumber \\
{\rm where} ~~~~\alpha (t) &=& \int^t J(t') dt', ~~~~{\rm and} ~~~~ \beta (t)
= \int^t \frac{1}{4}[\Delta(t')+2 B(t')] dt'. \label{mssceq2} \end{eqnarray}
Using (\ref{mssceq2}), one gets
\begin{eqnarray} f_{s r}(t) &=& \langle s | e^{-i \int^t H(t') dt'} |r\rangle
= \langle s | \phi(t) \rangle \nonumber \\
&=& \sum_k e^{-i \left[ k(r-s) + 2 \alpha (t) \cos(k) + \beta (t) \right]} .
\end{eqnarray}
For an infinite chain, the momentum sum can be converted to an integral
and exactly evaluated to yield
\begin{eqnarray} f_{sr}(t) &=& J_{r-s}(2\alpha (t)) ~e^{-i \beta (t)}.
\label{mssevo1} \end{eqnarray}
{}From this result, we note the following points. First, we need to
choose a time when we shall perform a measurement on the state. This
time, $t_0$, is chosen so as to maximize the fidelity of the state
transfer. In the present model, this occurs at the time $t_0$ when
the argument $2 \alpha (t_0)$ of the Bessel function approximately
equals $r-s$. This suggests that one can reach the maximum fidelity
(i.e., maximum $|f_{sr}(t)|$ and maximum $Re[f_{sr}(t)]$ which
requites a separate adjustment of the phase factor) for a given
separation $r-s$ at a much shorter time for a suitable non-adiabatic
dynamics. This ensures faster communication through the channel.
Note that by choosing an appropriate form of $J(t)$, the communication
can be made exponentially faster compared to adiabatic dynamics
since we may ramp up the effective instantaneous velocity so that a
given separation $r-s$ is reached at a much shorter time. Second,
the non-adiabatic dynamics gives us an additional handle on the
phase and hence the real part of $f_{sr}(t)$. Thus one can adjust
the phase using a user-chosen classical control parameter (such as
frequency in the case of AC dynamics) to obtain maximum fidelity for a
given $|f_{sr}(t)|$. Finally, it is straightforward to generalize the
derivation of $f_{sr}(t)$ to higher dimensions. The result for a 2D system is
\begin{eqnarray} f_{{\bf s r}} ~=~ J_{r_x-s_x}(2\alpha (t)) ~J_{r_y-s_y}
(2 \mu \alpha (t)) ~e^{-i \beta (t)}, \label{mss2deq} \end{eqnarray}
where $\mu$ is an anisotropy parameter which signifies the relative
strengths of couplings of the $S_x$ and $S_y$ terms in the two
orthogonal spatial directions. For $\mu=1$, i.e., the isotropic
case, we find that the fidelity is maximized when propagation takes
place along the diagonal. But in general, the angle of maximum propagation 
is a function of $\mu$ and this can in general also be controlled. A similar 
analysis can be easily extended to higher dimensions; however, as can be 
seen from (\ref{mss2deq}), the fidelity of the qubit transfer using this 
method rapidly decays with increasing dimensions.

Thus we find, via a simple analysis of a Heisenberg spin model with a
time-dependent Hamiltonian, that both the fidelity and the speed of
quantum communication may be improved by using suitable
non-equilibrium dynamics. We have also shown that such a procedure
can lead to direction specific state transfer in higher dimensional
spin systems. Since engineering such time-dependent Hamiltonians
have become an experimental reality, this might, in principle, provide a
realizable way for faster communication of qubits in future experiments.

\section{Discussion}

To summarize, we first discussed the response of a system of interacting
bosons in a 1D optical lattice to a sudden change in a harmonic trap
potential. The system can be mapped to a system of dipoles described by an
Ising order parameter. After the sudden shift, the order parameter oscillates
in time; the amplitude of oscillations depends on the initial and final
trap potentials. We then considered an infinite range ferromagnetic Ising
model in a transverse magnetic field; we studied what happens after the
field is changed suddenly. Once again, the variation of the order parameter
(the magnetization in this problem) with time depends in an interesting way
on the initial and final fields and the system size.

Next, we considered what happens when a system is taken across a quantum
critical point or a critical surface in a non-adiabatic way which is
governed by a quench time $\tau$. This leads to the production of defects;
the density of defects scales as an inverse power of $\tau$, where the
power depends on the dimensionalities of the system and the critical surface
and the critical exponents $z$ and $\nu$. This was illustrated by considering
the Kitaev model which is an exactly solvable spin-1/2 model defined on the
honeycomb lattice; this can be solved by mapping it to a system of
non-interacting Majorana fermions using a Jordan-Wigner transformation. We
then considered the effect of taking a system across a QCP in a non-linear
manner at time $t=0$; the non-linearity is parametrized by an exponent
$\alpha$. We found that two different things happen depending on whether the
system passes through the QCP at $t=0$ or at a non-zero value of $t$. In the
former case, the power appearing in the scaling of the defect density with
$\tau$ also depends on $\alpha$; in the latter case, the power is the same
as in a linear quench (corresponding to $\alpha =1$), but the effective
quench time $\tau_{\rm eff}$ depends on $\alpha$. These ideas are illustrated
by considering two models in 1D, namely, a 1D version of the Kitaev model,
and the Ising model in a transverse magnetic field; both of these can solved
by mapping them to systems of non-interacting fermions by a Jordan-Wigner
transformation. We then discussed some experimental systems where our
results for the defect scaling can be checked.

Finally, we used some Heisenberg spin-1/2 models in one and two dimensions
to discuss how a qubit can be transferred across the system. In particular,
we examined how the speed and fidelity of the transfer can be maximized
by choosing the couplings in the Hamiltonian appropriately.

Before ending, we would like to mention two possible extensions of the work
discussed here. First, it would be interesting to study whether the defects
studied in Sect. \ref{mssnon-adiabatic} have any non-trivial topology
associated with them. If they are not topological, it would be interesting
to find other ways of changing the parameters in the Hamiltonian in order to
produce defects which do have a topological character. It is known that
topology can affect defect production in a profound way \cite{mssbermudez}.
Secondly, the defect
density discussed in Sect. \ref{mssnon-adiabatic} follows from the
density matrix of a single site obtained by integrating out all the other
sites of the system. It is interesting to compute the two-site density
matrix and use that to obtain various measures of two-site entanglement. This
has been studied recently both for a sudden quench \cite{mssas2} and a slow
quench \cite{msssengupta09} through a QCP. Other kinds of entanglement
produced due by a quench have also been studied \cite{msscincio}.

\vskip .8 cm
\noindent {\large{\bf Acknowledgements}}
\vskip .4 cm

We thank Amit Dutta and Anatoli Polkovnikov for stimulating discussions.

\end{document}